\documentclass[prd,amsmath,amssymb,aps,amsfonts,notitlepage,superscriptaddress,eqsecnum,nofootinbib,10pt]{revtex4-1}
\usepackage{graphicx}       
\usepackage{dcolumn}        
\usepackage{bm}             
\usepackage{tensor}
\usepackage[utf8]{inputenc}
\usepackage[colorlinks]{hyperref}
\usepackage{color}
\usepackage[usenames,dvipsnames,svgnames,table]{xcolor}
\usepackage{float}
\usepackage{silence}
\allowdisplaybreaks[1]

\usepackage{dcolumn}
	\newcolumntype{.}{D{.}{.}{13}}
	\newcolumntype{d}[1]{D{.}{.}{#1}}
\usepackage{longtable}
\usepackage{subfigure}
\usepackage{rotating}
\usepackage[small,compact]{titlesec}
\usepackage{etoolbox}
\makeatletter
\patchcmd{\ttlh@hang}{\parindent\z@}{\parindent\z@\leavevmode}{}{}
\patchcmd{\ttlh@hang}{\noindent}{}{}{}
\makeatother
\usepackage{fancyhdr}
\usepackage{times}
\usepackage{multirow}
\usepackage{sidecap}
\usepackage{eurosym}
\usepackage{setspace}

\definecolor{CiteColor}{rgb}{0,0.5,0}
\hypersetup{citecolor=CiteColor}
\definecolor{RefColor}{rgb}{0.55,0,0}
\hypersetup{linkcolor=RefColor}
\definecolor{darkgreen}{rgb}{0.2,0.7,0.2}

\newcommand{\beq}{\begin{equation}}
\newcommand{\eeq}{\end{equation}}
\renewcommand{\c}{\text{circ}}
\newcommand{\e}{{e\to 0}}

\newcommand{\nn}{\nonumber}
\newcommand{\EE}{E}
\newcommand{\LL}{L}
\newcommand{\eb}{\bar{e}}
\newcommand{\uu}{\bar{u}}
\newcommand{\KK}{\mathcal{K}}

\newcommand{\ord}{\mathcal{O}}

\newcommand{\f}{\frac}
\newcommand{\pd}{\partial}
\newcommand{\bt}{\beta}

\newcommand{\al}{\alpha}

\newcommand{\be}{\begin{equation}}
\newcommand{\ee}{\end{equation}}
\newcommand{\ba}{\begin{eqnarray}}
\newcommand{\ea}{\end{eqnarray}}
\newcommand{\bi}{\begin{itemize}}
\newcommand{\ei}{\end{itemize}}
\newcommand{\bef}{\begin{frame}}
\newcommand{\ef}{\end{frame}}
\newcommand{\es}{\end{small}}
\newcommand{\Lim}[1]{\raisebox{0.5ex}{\scalebox{0.8}{$\displaystyle \lim_{#1}\;$}}}
\newcommand\T{\rule{0pt}{2.6ex}}       
\newcommand\B{\rule[-1.2ex]{0pt}{0pt}} 

\newcommand{\Delhat}{\hat{\delta}}
\newcommand{\vsp}{\vspace{0.2cm}}

\begin{document}

\title{Self-force correction to geodetic spin precession in Kerr spacetime}

\author{Sarp Akcay}
\affiliation{The Institute for Discovery, School of Mathematics \& Statistics, University College Dublin, Belfield, Dublin 4, Ireland.}


\begin{abstract}
We present an expression for the gravitational self-force correction to the geodetic spin precession of a spinning compact object with small, but non-negligible mass in a bound, equatorial orbit around a Kerr black hole. We consider only conservative back-reaction effects due to the mass of the compact object ($m_1$) thus neglecting 
the effects of its spin $s_1$ on its motion, i.e, we impose $s_1 \ll G m_1^2/c$ and $m_1 \ll m_2$, where $m_2$ is the mass
parameter of the background Kerr spacetime. We encapsulate the correction to the spin precession in $\psi$, the ratio of the accumulated spin-precession angle
to the total azimuthal angle over one radial orbit in the equatorial plane. Our formulation considers the gauge-invariant $\ord(m_1)$ part of the correction to
$\psi$, denoted by $\Delta\psi$, and is a generalization of the results of [Class. Quan. Grav., 34, 084001 (2017)] to Kerr spacetime. 
Additionally, we compute the zero-eccentricity limit of $\Delta\psi$ and show that this quantity differs from the circular orbit $\Delta\psi^\text{circ}$ by 
a gauge-invariant quantity containing the gravitational self-force correction to general relativistic periapsis advance in Kerr spacetime. Our result for $\Delta\psi$ is expressed in a manner that readily accommodates numerical/analytical self-force computations, e.g., in the radiation gauge, and paves the way for the computation of a new eccentric-orbit Kerr gauge invariant beyond the generalized redshift.
\end{abstract}
\maketitle

\section{Introduction}\label{Sec:Intro}
With three confirmed detections and a statistically significant trigger event, the third quarter of this decade marks the beginning of the era of gravitational-wave astronomy \cite{Abbott:2016blz, Abbott:2016nmj, TheLIGOScientific:2016pea, PhysRevLett.118.221101}. Given the inferred event rate of $\gtrsim 9\,\text{Gpc}^{-3}$ per year \cite{TheLIGOScientific:2016pea}, 2017 will surely herald more discoveries of gravitational radiation emitted by compact binary systems with potential detection of $\sim 100$ black hole mergers per year
 \cite{Belczynski:2016obo} when Advanced LIGO reaches its full design sensitivity by 2019 \cite{TheLIGOScientific:2014jea}.
While these detections were making headlines, the LISA Pathfinder spacecraft was quietly floating at the L1 point, maintaining test masses in near free fall (with sub-Femto-$g$ accelerations) in the relevant frequency range. 
Its performance exceeded its mission requirements by a factor of five thus meeting the proposed LISA mission noise budget \cite{Armano:2016bkm}.
The path to LISA's launch is now clear and an official mission proposal has been selected by the European Space Agency in January 2017 \cite{Audley:2017drz}.

LISA will be operating in the $\sim 10^{-4}$ - $1$ Hz frequency range. As such it will detect gravitational radiation from a myriad of sources
such as merging massive black holes, galactic white dwarf binaries and extreme mass ratio inspirals (EMRIs). 
The latter are especially important because they enable high-precision measurements of the first few multipole moments of the massive component of the binary \cite{Audley:2017drz}, which will provide tests of the ``Kerr-ness'' of these massive black holes \cite{Gair:2012nm} conjectured by
Ref.~\cite{Rees:1984si}. Furthermore, EMRI observations will provide information on the distribution of massive black hole masses and spins,
and estimations of their population at low redshift ($z\lesssim 1$) \cite{Gair:2010yu}. In addition, luminosity distances to EMRIs will be determined to 
\%1 precision \cite{AmaroSeoane:2012je} which may then enable the measurement of the Hubble constant to $\lesssim\%1$ \cite{PhysRevD.77.043512}.

EMRIs are systems in which a compact object (CO) of mass $m_1 \sim 1 - 50 M_\odot$ (a neutron star or a black hole) orbits a (super)massive black hole of
mass $m_2 \sim 10^5 - 10^7 M_\odot$. The CO's mass causes a radiation (back)reaction making it lose energy and angular momentum thereby driving a
gradual inward spiral which eventually ends with its plunge into the central black hole after $\sim q^{-1}$ orbits, where $q\equiv m_1/m_2 \ll 1$ is the mass ratio.
To achieve the desired accuracy for multipole-moment measurements, LISA needs to track the EMRI phase evolution to better than 0.1 radians over the
course of an inspiral \cite{Osburn:2015duj}. This imposes a total relative phase error of $ \ord(q^{-1})$ on numerical models of EMRIs which will be needed to
extract the EMRI signals from LISA noise \cite{Barack:2009ux}. Due to the extreme-mass ratios, numerical modelling of EMRIs is currently out of reach of numerical relativity (NR) and because of the small binary separations ($r\lesssim10 G m_2/c^2$), post-Newtonian (pN) theory is inadequate. However,
progress is being made in both fields to extend their coverage in the $r$-$q$ parameter space of compact binary systems \cite{Lousto:2010ut, Damour:2014jta, Damour:2015isa, Bernard:2016wrg}.

The gravitational self-force (GSF) approach which, at its core, is a perturbative treatment, is ideally suited to tackle the EMRI evolution challenge.
By introducing the CO into the background (unperturbed) spacetime order by order in $m_1$, the GSF approach successfully incorporates both the dissipative
and conservative effects of radiation reaction \cite{Poisson:2011nh, Barack:2009ux, Thornburg:2011qk, Pound:2015tma}. The GSF programme began in the mid 1990s
\cite{Mino:1996nk, Quinn:1996am} and combines a plethora of techniques including black hole perturbation theory \cite{Regge-Wheeler, Zerilli:1970,
 Teukolsky:1972my, Chrzanowski:1975wv}, matched asymptotic expansions \cite{Mino:1996nk, Pound:2009sm, Harte:2011ku} and rigorous regularization methods \cite{Barack:2001gx, Detweiler:2002mi, Vega:2007mc, Dolan:2010mt, Wardell:2015ada}. Thus far, all computations of the GSF have been linear in the mass ratio:
one solves the linearized Einstein field equation sourced by a term proportional to $m_1$. Although this is sufficient to capture the general characteristics
 of EMRI evolution \cite{Warburton:2011fk, Osburn:2015duj}, to meet the error requirements of the LISA mission, the $\ord(m_1^2)$ contribution to the GSF 
 will be required \cite{Hinderer:2008dm}. Significant progress has been made in this second-order-GSF sector in the last five years \cite{Detweiler:2011tt,Pound:2012prl, Gralla:2012prd,Pound:2014xva,Pound:2015wva,Miller:2016hjv,Pound:2017psq, Warburton:2013lea, Wardell:2015ada} and results that can
be compared with pN theory and NR are expected soon. 

The effects of radiation reaction on the motion of the CO can be viewed in two perspectives: (i) a self-forced motion where the CO is accelerated
away from the background geodesic worldline, or (ii) a geodesic motion in an effective perturbed spacetime with metric 
$g_{ab} = \bar{g}_{ab}+h^R_{ab}$, where $\bar{g}_{ab}$ is the metric of the background spacetime 
and $h^R_{ab}$ is a certain smooth vacuum
solution to the $\ord(m_1)$ perturbation equations whose solution $h_{ab}$ is decomposed into $h^R_{ab}$ and a singular piece $h^S_{ab}$ \cite{Detweiler:2002mi}. This procedure is known as the Detweiler-Whiting decomposition \cite{Detweiler:2000gt,Gralla:2008fg,Harte:2011ku}. 
Relatedly, one can also separate the GSF into dissipative and conservative pieces
under time reversal symmetry. The time-antisymmetric (dissipative) part causes energy and angular momentum decay in time whereas the time-symmetric 
(conservative) part shifts the orbital parameters \cite{Barack:2009ey,Barack:2011ed}.

In 2008 Detweiler showed that the $\ord(m_1)$ \emph{conservative} shift to the time component of the four-velocity (for circular timelike geodesics in the perturbed spacetime) is invariant under $\ord(m_1)$ gauge transformations that respect the helical symmetry of circular geodesics in the background Schwarzschild spacetime \cite{Detweiler:2008ft}. Given by $\Delta u^t \equiv \bar{u}^t h^{R,\text{cons}}_{ab}\bar{u}^a\bar{u}^b/2$, this quantity has been dubbed ``the redshift invariant'', where ${}^\text{cons}$ denotes the conservative part.
Following Detweiler's proof, 
two numerical computations of $\Delta u^t$ performed in different gauges were shown to agree \cite{Sago:2008id}, and further concordance was subsequently obtained in a third gauge \cite{Shah:2010bi, Shah:2013uya}. Soon after these initial perturbation-theory approaches, the redshift invariant was computed in both pN theory
\cite{Blanchet:2009sd, Blanchet:2010zd, Blanchet:2013txa} and numerical relativity \cite {Zimmerman:2016ajr} with excellent agreement within overlapping
domains between the GSF approach, pN theory and NR. Moreover, a functional relationship was obtained between the binding energy of a non-spinning binary
and $\Delta u^t$ using the first law of binary black hole mechanics \cite{LeTiec:2011dp,LeTiec:2011ab,Blanchet:2012at}.
The circular-orbit redshift invariant is now known to very high pN order in Schwarzschild spacetime \cite{Bini:2015bla, Johnson-McDaniel:2015vva,
  Kavanagh:2015lva, Shah:2015nva}.
  
Other invariants in Schwarzschild spacetime were soon identified such as the $\ord(m_1)$ shifts to the frequency of the innermost stable circular orbit (ISCO) \cite{Barack:2009ey} and to the general relativistic periapsis advance \cite{Barack:2011ed}. Using these GSF results, Damour and collaborators determined the 
strong-field behavior of the $\ord(m_1)$ part of the effective-one-body (EOB) potentials \cite{Damour:2009sm, Barack:2010ny, Akcay:2012ea} which
were then used to numerically compute the unknown higher-order coefficients in pN series expansions of these potentials.
The pioneering works of Refs.~\cite{Sago:2008id,Shah:2010bi,Blanchet:2009sd} along with the GSF-EOB collaborations marked the emergence of
a new field of synergistic studies --- among the various approaches to treat the two-body problem in GR --- which are based on cross-cultural comparisons of
gauge-invariant quantities. Landmark studies on the binary binding energy \cite{LeTiec:2011dp} and the periapsis advance for non-spinning \cite{LeTiec:2011bk}
and spinning binaries \cite{PhysRevD.88.084005} illustrate the power of these synergies.

The classification of the above $\ord(m_1)$ invariants for circular orbits in Schwarzschild spacetime naturally led to three fronts for progression:
(i) finding invariants that come from first and higher-order derivatives of $h_{ab}^R$ in Schwarzschild and Kerr spacetimes, 
(ii) generalizing the circular-orbit formulation to generic bound orbits in Schwarzschild spacetime, and then to
(iii) bound orbits (eccentric and inclined) in Kerr spacetime. In the last five years, all of these challenges
have been taken up by the GSF community with successful results. Dolan and collaborators systematically constructed higher order-derivative invariants for
circular orbits in Schwarzschild and Kerr spacetimes starting with the geodetic spin-precession invariant at $n=1$ \cite{Dolan:2013roa, Bini:2014ica, Dolan:2014pja, Bini:2015mza, Shah:2015nva}, the tidal eigenvalues at $n=2$ \cite{Dolan:2014pja, Bini:2014zxa, Bini:2015kja} and the octupolar tidal invariants at $n=3$ \cite{Bini:2014zxa, Bini:2015kja, Nolan:2015vpa}, where $n$ denotes the order of the highest derivative of $h^R_{ab}$. On front (ii), a generalized redshift invariant was computed for eccentric orbits in Schwarzschild
spacetime 
\cite{Barack:2011ed} and subsequently shown to agree with 3-pN accurate expressions \cite{Akcay:2015pza}. Then, following the methodology of Ref.~\cite{Barack:2011ed}, the generalized spin-precession invariant was obtained in Schwarzschild using both the GSF approach and pN theory \cite{Akcay:2016dku}. 
As for front (iii), the GSF computation of the redshift invariant for eccentric equatorial orbits \cite{vandeMeent:2015lxa} 
and the periapsis advance for nearly circular orbits in Kerr \cite{vandeMeent:2016hel} have been the most recent advances. 
As there are currently no results for $h_{ab}^R$ and/or the GSF along fully generic (\emph{inclined} and eccentric) orbits in Kerr, no invariants have been 
computed, but there are ongoing efforts.

Our work in this article is a new contribution to front (iii). We present an expression for the gauge-invariant $\ord(m_1)$ contribution to
the geodetic spin precession of a spinning CO (e.g., a small Kerr black hole, or a gyroscope) in an eccentric, \emph{equatorial} orbit around a Kerr black hole
with mass $m_2$ and adimensionalized spin $a= s_2 c/(G m_2^2)$. As we compute the back-reaction effects due to the mass of the CO, but not its spin,
we impose the condition $s_1 \ll G m_1^2/c$ along with the condition $m_1\ll m_2$ necessary for linear perturbation theory. We give a general expression
for this invariant quantity, which we denote by $\Delta\psi$, in Eqs.~(\ref{eq:Delta_psi1}, \ref{eq:Delta_Psi_formula}) with each term expanded in detail
in the remainder of Sec.~\ref{Sec:GSF}. For eccentric, equatorial orbits in Kerr, $\psi$ measures the net fractional precession with respect to the 
azimuthal phase $\Phi$ accumulated over one radial orbit, i.e.,
\be
\psi = \f{\Phi - \Psi}{\Phi}\, , \label{eq:psi_general}
\ee
where $\Psi$ is the total spin-precession angle (in the equatorial plane) over one radial orbit. As equatorial orbits in Kerr remain equatorial (neglecting
spin-spin interactions), $\Psi$ can be quantified in terms of the rotation of the equatorial ``legs'' of an orthonormal tetrad with respect to a preferred 
basis.

Our contributions here are twofold in the sense that by extending the eccentric-orbit $\Delta\psi$ formulation of Ref.~\cite{Akcay:2016dku} from Schwarzschild to Kerr background we are also introducing the first $\ord(m_1)$ invariant quantity beyond the Detweiler redshift for the eccentric Kerr case.
Our final expressions making up $\Delta\psi$ can readily accommodate numerical GSF results that will be obtained via the approach of Ref.~\cite{vandeMeent:2016hel} or similar techniques. We leave this for future work.

This article is organized as follows. In Sec.~\ref{sec:test_body}, we review equatorial geodesic motion and spin precession of test masses in Kerr spacetime.
In Sec.~\ref{Sec:GSF}, we derive an expression for $\Delta\psi$ using the GSF formalism applied to the case of a spinning CO in an eccentric, equatorial orbit around a Kerr black hole.
In Sec.~\ref{sec:e_to_limit} we present a detailed calculation of the zero-eccentricity limit of $\Delta\psi$ and show how it is related to the circular-orbit
$\Delta\psi^\c$ of Ref.~\cite{Dolan:2014pja} via the another gauge invariant: the $\ord(m_1)$ correction to the periapsis
advance in the eccentricity $\to 0$ limit. We conclude with a summary of our results and a discussion of near-future plans in Sec.~\ref{Sec:discussion}. 

We follow the theorist's convention of natural units, i.e., $G=c=1$, and the relativist's convention of $(-,+,+,+)$ signature for the spacetime metric.
Furthermore, we rescale all physical quantities in terms of $m_2$ which we set equal to 1. As a result, all the $\ord(m_1)$ quantities mentioned thus far will
become $\ord(q)$ quantities with $q\ll 1$. This is standard practice in the self-force literature. If necessary, the proper units and dimensions can be recovered straightforwardly.
We employ the Latin letters $a,b,c, \ldots $ to denote spacetime indices and $i,j,k, \ldots$ for spatial indices. The Greek letters $\alpha,\beta,\ldots$ represent tetrad indices.
Square brackets $[\ldots ]$ denote antisymmetrization while
$(\ldots )$ denote symmetrization over indices within the 
brackets (parentheses).

\section{Geodesics and spin precession in the test-body limit\label{sec:test_body}}
For the remainder of this article, we follow the notation of Ref.~\cite{Akcay:2016dku} (henceforth Paper I) and borrow from their discussion.

\subsection{Geodetic spin precession}\label{Sec:geodetic_precession}\vsp
Let us begin by considering a ``test'' gyroscope of negligible mass with spin four-vector $s_1^a$ in a bound geodesic trajectory $z^a(\tau)$ in Kerr spacetime. The gyroscope's unit timelike four-velocity is given by $u^a = dz^a/d\tau$ which is naturally parallel-transported, i.e., $Du^a/d\tau\equiv u^b \nabla_b u^a=0$, where $\nabla_a$ is the covariant derivative compatible with the metric of the spacetime $g_{ab}$. 
In the rest frame of the gyroscope, $s^a_1$ only has spatial components; thus $s_{1a} u^a=0$. 
The spin vector is also parallel-transported \cite{Harte:2011ku}, which implies that $s_1^2=g_{ab}s_1^a s_1^b$ is conserved.

We may `attach' a tetrad $e^a_\alpha$ to the gyroscope by setting $e^a_0=u^a$, which is orthonormal in the usual sense: $g_{ab}e^a_\alpha e^b_\beta=\eta_{\alpha\beta} = \text{diag}[-1,1,1,1]$. Projecting the spin vector into spatial ($i=1,2,3$) components via $s_i = e^a_i s_a$ allows us to write
the parallel-transport equation $D s_{1a}/d\tau=0$ as
\be
\f{ds_{1i}}{d\tau} = \epsilon_{ijk}\omega^j s_1^k\, , \label{eq:dsdtau}
\ee
where $\omega_i \equiv -\tfrac{1}{2}\epsilon_{ijk}\omega^{jk}$, $\epsilon_{ijk}$ is the Levi-Civita symbol, and
\be
\omega_{ij} \equiv g_{ab}\, e^a_i \f{De_j^b}{d\tau} = -\omega_{ji}\, .\label{eq:omega_ij}
\ee
Eq.~(\ref{eq:dsdtau}) is a precession equation for the parallel-transported spin vector with respect to a reference basis and $|\omega^i|$ is the 
proper-time precession frequency. Note that $\omega_i$ depends on the choice of the reference basis; therefore if a parallel-transported tetrad is chosen 
for $e^a_i$, then $\omega^i = 0$. Following Paper I, we will adopt Marck's tetrad as our basis as explained in Sec.~\ref{Sec:Equatorial_geodesics_in_Kerr}.

Substituting Eq.~(\ref{eq:omega_ij}) into Eq.~(\ref{eq:dsdtau}) simplifies the latter to
\be
\f{ds_{1i}}{d\tau} = \omega_{ij} s_1^j\, . \label{eq:dsdtau2}
\ee
$\omega_{ij}$ is a rank-2 antisymmetric tensor of dimension 3 so it has only three nonzero, independent components: $\omega_{12},\omega_{13}$ and $\omega_{23}$.
Our preferred basis has the property that $e^a_2 \propto \tfrac{\pd}{\pd\theta}^a$ (orthogonal to the equatorial plane) so $e_2^a$ is parallel-transported
along the geodesic. This leads to $\omega_{12}=\omega_{23}=0$, implying that the spin vector evolves (precesses) in the 1-3 plane only with proper-time frequency $\omega_{13}$. 
We can immediately solve Eq.~(\ref{eq:dsdtau2}) to obtain
\be
s^{j=1}_1 + i s^{j=3}_1 = S_{||} \exp\left(i\int^\tau \omega_{13}(\tau)d\tau \right), \quad s^{j=2}_1 = S_\perp,
\ee
where we introduced components of $s_{1}$ parallel ($S_{||}\in\mathbb{C}$) and perpendicular ($S_\perp\in\mathbb{R}$)
to the orbital plane with the condition $s_1^2= g_{ab} s_1^a s_1^b= |S_{||}|^2+S_\perp^2$. Note that this $||, \perp$ labelling is the opposite of Paper I's. We can now define the 
total accumulated geodetic precession over one radial period
\be
\Psi = \int_0^\mathcal{T} \omega_{13}(\tau)\, d\tau = \int_0^T \f{\omega_{13}(t)}{u^t} \,dt, \label{eq:Psi_general}
\ee
where $\mathcal{T}, T$ are the radial periods with respect to proper and coordinate times $\tau,t$, respectively, and $u^t = dt/d\tau$.

As explained in Paper I, eccentric orbits in Schwarzschild spacetime have a discrete isometry with respect to the radial period $T$. 
In Kerr, this discrete isometry still exists for equatorial orbits. Therefore, if we restrict our
attention to triads $e^a_i$ that rotate once in passing through $2\pi$ around the black hole in the $\phi$-direction (like the spherical polar basis) 
$\Psi$ becomes insensitive to a specific choice of reference basis within a general class that respects the discrete isometry. 
A detailed discussion of this argument can be found in Sec.~II.B of Paper I.

\subsection{Equatorial timelike geodesics in Kerr spacetime}\label{Sec:Kerr_geodesics}
Immediately confining the motion to the equatorial plane by setting $\theta = \pi/2$, we obtain,
in standard Boyer-Lindquist (BL) coordinates $\{t,r,\theta,\phi\}$, the equatorial Kerr line element
\be
 ds^2= g_{ab}\, dx^a dx^b = -\left(1-\frac{2 m_2}{r}\right)dt^2 - \frac{4 m_2\, a}{r} dt d\phi + \f{r^2}{\Delta}dr^2 + \left(r^2+a^2+\f{2m_2\, a^2}{r}\right)d\phi^2 \label{eq:Kerr_eq_line_element},
 \ee
where $\Delta = r^2 -2 m_2 r +a^2$. The CO follows a timelike geodesic trajectory whose spatial projection remains in the equatorial plane
and whose tangent vector is given by the four-velocity $u^a = [\dot{t}\,,\dot{r}\,,0,\dot{\phi}\,]^T$, where the overdot denotes the derivative with respect to proper time $\tau$.
The conserved energy $\EE$ and angular momentum $\LL$ of the CO are related to the components of $u^a$ via
\begin{align}
 \EE &= \left(1-\f{2 m_2}{r}\right) \dot{t} + \f{2 m_2}{r} \dot{\phi}, \label{eq:Energy_tdot_phidot}\\
 \LL &= -\f{2 m_2}{r}\dot{t} + \left(r^2+a^2+\f{2m_2\, a^2}{r}\right) \dot{\phi}\label{eq:AngMom_tdot_phidot}.
\end{align}
Following Ref.~\cite{Glampedakis:2002ya}, we define $x\equiv \LL - a \EE$ and ${\mathfrak T} \equiv \EE(r^2+a^2)-a \LL= \EE r^2-ax$; then using Eqs.~(\ref{eq:Energy_tdot_phidot}, 
\ref{eq:AngMom_tdot_phidot}) we obtain
\begin{align}
 \dot{t} &= \f{(r^2+a^2)\,{\mathfrak T}}{r^2\Delta} +\f{ax}{r^2}, \qquad \dot{\phi} = \f{a {\mathfrak T}}{r^2\Delta}+\f{x}{r^2}\, .\label{eq:tdot_phidot}
\end{align}
For the radial equation, we have the following well-known expression
\be
 \dot{r} \equiv R(r)= \pm \f{1}{r^2}\left[{\mathfrak T}^2-\Delta(r^2+x^2)\right]^{1/2} \label{eq:rdot}.
 \ee
Bound orbits in the equatorial plane have well-defined points of closest (periapsis) and farthest (apoapsis) approach denoted by $r_\text{min}$ and $ r_\text{max} $, 
respectively. One can parametrize a bound, equatorial geodesic by these two parameters or by the dimensionless semi-latus rectum $p$ and eccentricity $e$, which
relate to $r_\text{min}, r_\text{max}$ via
\be
r_\text{min} = \f{m_2\, p}{1+e},\quad r_\text{max} = \f{m_2\, p}{1-e} \label{eqs:r_min_r_max}.
\ee
The radial geodesic equation can be solved analytically at $R(r_\text{min}) = R(r_\text{max})=0$ to yield
\begin{align}
\EE &= \left[ 1 -\f{m_2}{p}(1-e^2)\left\{1-\f{x^2}{p^2}(1-e^2)\right\}\right]^{1/2}, \label{eq:Energy_p_e}\\
\LL &= a \EE +x \label{eq:AngMom_p_e},
\end{align}
where $x^2$ is an analytic function of $p,e$, and $a$ that satisfies the quartic equation \cite{Glampedakis:2002ya}
\be
F(p,e) x^4 + N(p,e) x^2 + C(p,e) = 0\, 
\ee
the solution to which is given by
\be
x^2 = \frac{-N \mp \Delta_x^{1/2}}{2 F}\, ,
\ee
where
\begin{align}
 F(p,e) &= \f{1}{p^3} \left[p^3-2m_2(3+e^2)p^2+m_2^2(3+e^2)^2p- 4 m_2 a^2(1-e^2)^2\right],\\
 N(p,e) &= \f{2}{p}\left[-m_2 p^2 + \left(m_2^2(3+e^2)-a^2\right)p - m_2 a^2(1+3e^2)\right],\\
 \Delta_x(p,e) &= \f{16 a^2m_2}{p^3}\left[p^4-4m_2p^3+2\left\{2m_2^2(1-e^2)+a^2(1+e^2)\right\}p^2-4m_2 a^2(1-e^2)p+a^4(1-e^2)^2\right]\, .
\end{align}
%
The sign of $a$ determines whether the orbit is prograde ($a>0$) or retrograde ($a<0$).

By introducing a relativistic anomaly $\chi \in [0,2\pi]$, we can parametrize the radial coordinate of the equatorial orbit as follows
\be
r(\chi) = \frac{p\, m_2}{1 + e \cos \chi}\, .  \label{eq:rchi}
\ee
Then using
\be
\f{d\tau}{d\chi} = \f{r^2 }{\sqrt{V_r}}, \label{eq:dtaudchi_p_e_chi}
\ee
we obtain
\begin{align}
 \dot{r}\ \ &=\f{e\sin\chi}{p}\sqrt{V_r}, \label{eq:rdot_p_e_chi}\\
 \f{dt}{d\chi} &= \f{V_t}{J \sqrt{V_r}} \label{eq:dtdchi_p_e_chi},\\
 \f{d\phi}{d\chi} &= \f{V_\phi}{J \sqrt{V_r}} \label{eq:dphidchi_p_e_chi},
\end{align}
where %
\begin{align}
V_r &\equiv x^2 + a^2 + 2x a E - \f{2m_2\, x^2}{p}(3+e\cos\chi) \label{eq:Vr},\\
V_t &\equiv a^2\EE-\f{2 a m_2\,x}{p}(1+e\cos\chi) + \f{\EE\, p^2}{(1+e\cos\chi)^2} \label{eq:V_t},\\
V_\phi &\equiv x+a\EE-\f{2 m_2\,x}{p}(1+e\cos\chi) \label{eq:V_phi},\\
J &\equiv 1-\f{2m_2}{p}(1+e\cos\chi) +\f{a^2}{p^2}(1+e\cos\chi)^2 . \label{eq:J}
\end{align}
We can now determine the $\{$proper, coordinate$\}$-time periods $\{\mathcal{T},T\}$, and the accumulated azimuthal angle $\Phi$ 
by integrating over one radial orbit parametrized by $\chi$, i.e., $\{\mathcal{T},T,\Phi\} = \int_0^{2\pi} \frac{d}{d \chi}\{\tau,t,\phi\} d\chi . $
These orbit integrals can also be written in terms of elliptic integrals \`{a} la Fujita and Hikida \cite{Fujita:2009bp}.\vspace{0.1cm}\\
Next, we introduce the radial and azimuthal frequencies with respect to coordinate time $t$
\be
\Omega_r = \f{2\pi}{T},\qquad \Omega_\phi = \f{\Phi}{T} \label{eq:Omega_r_Omega_phi}\, .
\ee
$\{\Omega_r, \Omega_\phi\}$ are quantities that can be measured by an observer at infinity (asymptotically flat spacetime), hence they provide a
gauge-invariant (``physical'') way to parametrize eccentric orbits as opposed to $\{p,e\}$ or $\{r_\text{min},r_\text{max}\}$ which are all gauge-dependent 
quantities (the radial coordinate $r$ is not gauge invariant \cite{Barack:2001ph}).

\subsection{Geodetic spin precession for equatorial geodesics in Kerr spacetime}\label{Sec:Equatorial_geodesics_in_Kerr} \vspace{0.1cm}
For $\theta = \pi/2$ the precession frequency reduces to
\beq
\omega_{13} = \frac{\sqrt{\KK}}{r^2 + \KK} \left( \EE + \frac{a}{\LL - a\EE} \right)  \label{eq:omKerr}.
\eeq
$\KK = (\LL -a\EE)^2$ is the Carter's constant for equatorial geodesics in Kerr.
This gives
\be
  \frac{d \Psi}{d \chi}= \omega_{13} \frac{d\tau}{d\chi} = \f{r^2\sqrt{\KK}}{r^2+\KK}\left(\EE+\f{a}{x}\right)\f{1}{\sqrt{V_r}} \label{eq:Psidot_p_e_chi}
\ee
with which we can immediately obtain the accumulated geodetic precession from Eq.~(\ref{eq:Psi_general})
\be
\Psi = \int_0^{2\pi} \f{d\Psi}{d\chi} \, d\chi \, . \label{eq:Psi_general2}
\ee
%
For our standard reference basis, we choose Marck's tetrad from Ref.~\cite{Marck:1983} 
originally presented in the canonical orthonormal basis $ds^2= \eta_{(A)(B)} \omega^{(A)} \omega^{(B)}$ ,
where $\omega^{(A)} = \left(e_0^{(A)},e_1^{(A)},e_2^{(A)},e_3^{(A)} \right) $
and\footnote{Note that Marck's expression  for $e^{(A)}_3$ given in Eq.~(67) of Ref.~\cite{Marck:1983} is missing the square root over $r^2+\mathcal{K}$ for the $(A)=\phi$ component.} 
\begin{align}
e_0^{(A)} &= u^{(A)} = \left[\f{{\mathfrak T}}{r\sqrt{\Delta}}, \f{r\dot{r}}{\sqrt{\Delta}}, 0, \f{a\EE-\LL}{r} \right]^T , & 
e_2^{(A)} &= \left[0, 0, \f{\LL-a\EE}{\sqrt{\mathcal{K}}}, 0 \right]^T ,  \label{eq:tetrad}   \\
e_1^{(A)} &= \left[\f{\dot{r}r^2}{\sqrt{\Delta(r^2+\mathcal{K})}}, \f{{\mathfrak T}}{\sqrt{\Delta(r^2+\mathcal{K})}}, 0, 0 \right]^T , & 
e_3^{(A)} &= \left[\sqrt{\f{\mathcal{K}}{\Delta(r^2+\mathcal{K})}}\,\f{{\mathfrak T}}{r},\sqrt{\f{\mathcal{K}}{\Delta(r^2+\mathcal{K})}}\,r\dot{r}, 0,\sqrt{\f{r^2+\mathcal{K}}{\mathcal{K}}}\,\f{a\EE-\LL}{r} \right]^T \nn.
\end{align}
Eq.~(\ref{eq:rdot_p_e_chi}) yields 
\beq
\dot{r}^2 = (\EE^2-1)+ \f{2m_2}{r} - \frac{1}{r^2}(x^2+2 x a E+a^2) +\f{2m_2\, x^2}{r^{3}}. \label{eq:energy_equation}
\eeq 
We can transform Marck's tetrad given above in the canonical basis to the BL coordinate basis 
$[\f{\pd}{\pd t}, \f{\pd}{\pd r}, \f{\pd}{\pd \theta}, \f{\pd}{\pd \phi}]^T$ via $\omega_{(A)}^a \equiv \left[ \omega^{(A)}_a\right]^{-1}$ to obtain
\begin{align}
 e^a_0 &= u^a =[ \dot{t}\,, \dot{r}\,, 0, \dot{\phi}\,]^T , \label{eq:e0} \\
 e^a_1 &= \left[\f{\dot{r}\, r (r^2+a^2)}{\Delta}, \f{\EE r^2-ax}{r},0,\f{a\dot{r}\,r}{\Delta}\right]^T, \label{eq:e1} \\
 e^a_3 &= \left[\alpha \f{(r^2+a^2)\,{\mathfrak T}}{r^2\Delta}+\f{ax}{\alpha r^2},\,\alpha \dot{r},\,0,\, \f{\alpha a {\mathfrak T}}{r^2\Delta}+\f{x}{\alpha r^2} \right]^T \label{eq:e3},\\
 e^a_2 &= [0,0,\f{1}{r},0]^T,
\end{align}
where $\alpha = \sqrt{\KK/(r^2+\KK)}$. It is straightforward to check that these form an orthonormal basis with respect to $g_{ab}$ i.e. 
$g_{ab}\, e^a_{\al}e^b_{\bt} = \eta_{\al\bt}$. The exact form of $e_2^a$ is irrelevant for our work, but can be obtained using the basis transformation. This orthonormal tetrad 
has the desired properties discussed in Sec.~\ref{Sec:geodetic_precession}
and reduces to the reference basis of Paper I for $a=0$ when normalized.
\section{Gravitational self-force method}\label{Sec:GSF} \vsp
We now consider our spinning CO to be massive, albeit much less than the central black hole, i.e. $m_1\ll m_2$.
We wish to calculate the effect of the small mass on the geodetic spin precession. More precisely, we want to establish a gauge-invariant 
relation between the $\ord(q)$ correction to $\psi$ and the invariant (observable) frequencies $\Omega_r, \Omega_\phi$ \cite{Barack:2011ed}. 
Our starting point is the assumption that there exists a well-defined function $\psi( \Omega_r,  \Omega_\phi, q)$, for any mass ratio $q$.
We isolate the contribution due to the $\ord(q)$ part of the back-reaction by definining the following operator
\beq
\Delta \psi(\Omega_r,  \Omega_\phi) \equiv  \left[ \psi( \Omega_r,\Omega_\phi, q) - \psi(\Omega_r, \Omega_\phi, 0) \right]_{\ord(q)},  \label{eq:Deltapsi}
\eeq
where the brackets denote the $\ord(q)$ part. 
The first crucial step in our approach is to ``turn off'' dissipation, i.e., to consider only the time-symmetric (conservative) part of $h_{ab}^R$ and the GSF $F^a$.
Then, invoking the perturbed geodesic interpretation of Detweiler and Whiting, there should still exist bound, equatorial geodesics in the perturbed spacetime $g_{ab} = \bar{g}_{ab}+h^{R,\text{cons}}_{ab}$, whose spatial orbits are ``closed''
in the sense of returning to the same radial position $r=r_p$ over a radial period $T$: $r_p(t+T)=r_p(t)$. Note that in this perturbed spacetime all quantities of interest now contain $\ord(q)$ contributions; thus we have introduced the overbar notation to denote unperturbed quantities.

As Eq.~(\ref{eq:Deltapsi}) indicates, we aim to compare $\psi$ in the perturbed and unperturbed spacetimes and compute the difference. 
This subtraction is only meaningful if certain quantities are held fixed when going from the unperturbed to the perturbed spacetime. 
In other words, we must pick a background reference geodesic
to compare with the perturbed-spacetime geodesic which has proper time $\tau$, coordinates $z^a(\tau)$ and orbital parameters $\{p,e,\chi\}$.
In Paper I, we had listed three different possible choices for this reference worldline and settled on the one that that has the same $\{p,e,\chi\}$
as the perturbed worldline. More specifically, we had set $p=\bar{p}, e=\bar{e}$, and $ \chi=\bar{\chi}$. Here, we follow suit.
Therefore, introducing the operator $\delta$ to denote the $\ord(q)$ difference between a quantity on a perturbed and unperturbed geodesic with the same $\{p,e,\chi\}$,
we immediately have that $\delta r = r(p,e,\chi) -\bar{r}(p,e,\chi) = 0$ since $r$ is only a function of $p,e$, and implicitly $\chi$, and
by definition $\delta p =\delta e =\delta\chi = 0$. However, these geodesics do not have the same $t$ and $\phi$ coordinates as we show explicitly in Sec.~\ref{Sec:Formulation_GSF_method}. This ``fixing'' of $\{p,e,\chi\}$ is explained in depth in Ref.~\cite{Barack:2011ed}.
Note that the $\delta$ perturbation of a physical quantity does not, in general, return a gauge-invariant quantity (e.g., $\delta\Omega_r\ne 0, \delta\Omega_\phi\ne 0$).
However, as these two frequencies are true observables, we must have that $\Delta\Omega_r = \Delta\Omega_\phi = 0$. 
This was shown in Ref.~\cite{Barack:2011ed} and has been confirmed numerically as well (cf. Refs.~\cite{Akcay:2015pjz, Akcay:2016dku}).

As shown in Paper I, if a certain background quantity $\bar{Y}$ is given in terms of the orbit integral of $\bar{y}\equiv d\bar{Y}/d\bar\tau$, i.e.,
\be
\bar{Y} = \int_0^{2\pi} \bar{y}\f{d\bar\tau}{d\chi}\, d\chi \label{eq:Ybar}
\ee
then $\delta Y$ is given by
\be
\delta Y = \int_0^{2 \pi} \left( \frac{\delta y}{\bar{y}} - \frac{\delta u^r}{\bar{u}^r} \right) \bar{y}\, \frac{d \bar{\tau}}{d\chi}\, d\chi \label{eq:delta_Y_generic},
\ee
where the second term arises from perturbing the proper time $\tau$ which is not fixed.
Hence for $y\in \{1,u^t,u^\phi,\omega_{13}\}$ we now have a well-defined algorithm to compute $\delta\{\mathcal{T},T,\Phi,\Psi\}$.
In Ref.~\cite{Barack:2011ed}, Barack and Sago showed how one may obtain $\Delta Y$ from $\delta Y$ by accounting for the facts that $\delta p =\delta e= \Delta\Omega_r = \Delta \Omega_\phi = 0$, but $\{\Delta p,\Delta e, \delta\Omega_r,\delta\Omega_\phi\} \ne 0$. This results in
\beq
\Delta Y = \delta Y - \f{\pd \bar{Y}}{\pd \bar\Omega_r}\delta \Omega_r - \f{\pd \bar{Y}}{\pd \bar\Omega_\phi}\delta\Omega_\phi \, .  \label{eq:Delfromdel}
\eeq
Applying this operator to $\psi$ we immediately obtain
\be
\Delta\psi = -\frac{\Delta\Psi}{\bar\Phi}, \label{eq:Delta_psi1}
\ee
where we used $\Delta\Phi= \Delta(\Omega_\phi/\Omega_r)=0$ and the numerator is given by
\be
\Delta\Psi = \delta\Psi -\f{\pd \bar\Psi}{\pd \bar\Omega_r}\delta \Omega_r - \f{\pd \bar\Psi}{\pd \bar\Omega_\phi}\delta\Omega_\phi = \delta\Psi -\f{\pd \bar\Psi}{\pd \bar{T}}\delta T - \f{\pd \bar\Psi}{\pd \bar\Phi}\delta\Phi, \label{eq:Delta_Psi_formula}
\ee
where we used the relations $\bar{T} = 2\pi/\bar\Omega_r, \bar\Phi = 2\pi \bar\Omega_\phi/\bar\Omega_r$.
The partials $\pd \bar\Psi/\pd\{\bar\Omega_r,\bar\Omega_\phi,\bar{T},\bar\Phi\}$ can be obtained in a straightforward fashion as outlined in Sec.~III.A of Paper I. We present the $e\to 0$ limits of these partial derivatives in App.~\ref{sec:AppA}. $\delta\Omega_{r,\phi}$ can be obtained
immediately from $\Omega_r=2\pi/T$ and $\Omega_\phi = \Phi/T$
using Eq.~(\ref{eq:delta_Y_generic}) with $Y=\{T,\Phi\}$ and
$y=\{u^t,u^\phi\}$.

\subsection{Formulation}\label{Sec:Formulation_GSF_method}
Our main task is then to compute $\delta\Psi$. From Eqs.~(\ref{eq:Psidot_p_e_chi}, \ref{eq:delta_Y_generic}) it is clear that this requires
$\delta\omega_{13}=\delta\left( g_{ab} e^a_3 \f{D e_1^b}{d\tau}\right) = -\delta\left(  g_{ab} e^a_1 \f{D e_3^b}{d\tau}\right)$. 
We commence by considering an orthonormal tetrad normalized with respect to the perturbed, dissipation-free spacetime $\bar{g}_{ab} + h^{R,\text{cons}}_{ab}$. 
Henceforth suppressing the superscript ``$R,\text{cons}$'' we perturb the tetrad as follows
\begin{align}
\delta {e}_0^a &=\delta u^a =  c_{00}u^a  +c_{01} e_1^a + c_{03} e_3^a , \nn \\
\delta e_1^a &= c_{10} u^a  +c_{11} e_1^a + c_{13} e_3^a , \nn \\
\delta e_3^a &= c_{30} u^a  +c_{31} e_1^a + c_{33} e_3^a .  \label{eq:eperturbed}
\end{align}
Imposing orthonormality conditions with $\ord(q^{\ge2})$ terms neglected in $ \eta_{\alpha\beta}=g_{ab}e^a_\alpha e^b_\beta \approx (\bar{g}_{ab}+h_{ab}) (\bar{e}^a_\alpha+\delta e^a_\alpha)(\bar{e}^a_\bt+\delta e^a_\bt)$ leads to
\begin{align}
c_{00} &= \frac{1}{2} h_{00} &
c_{11} &= -\frac{1}{2} h_{11}, &
c_{33} &= -\frac{1}{2} h_{33}, \nn \\
c_{10} &= h_{01}  + c_{01}, &
c_{30} &= h_{03}  + c_{03}, &
c_{13} + c_{31} &= -h_{13} , \label{eq:cco}
\end{align}
where
\beq
h_{\alpha \beta} \equiv h_{ab} e_\alpha^a e_\beta^b .
\eeq
The tangent vector $u^a = e_0^a$ is written as
\be
u^a = \left[\dot{t}+\delta\dot{t}, \dot{r}+\delta\dot{r},\, 0,\, \dot{\phi}+\delta\dot{\phi}\right]^T\, , \label{eq:uperturbed}
\ee
where $\dot{t},\dot{\phi}$, and $\dot{r}$ are the background quantities of Eqs.~(\ref{eq:tdot_phidot}, \ref{eq:rdot_p_e_chi}). 
Then, the relation $u_a u^a=-1$ allows us to write the $\ord(q)$ parts in the following manner
\be
\delta\{\dot{t},\dot{r},\dot{\phi}\} = \f{h_{00}}{2}\{\dot{t},\dot{r},\dot{\phi}\} + \hat{\delta}\{\dot{t},\dot{r},\dot{\phi}\} \label{eq:delta_tdot_rdot_phidot} \, ,
\ee
where the linear operator $\hat\delta$ acts on a generic quantity $Y(p,e,\chi,\EE,\LL)$ as follows:
\be
\hat\delta Y = \left.\f{\pd Y}{\pd E}\right|_{\bar{\EE},\bar{\LL}}\,\hat\delta\EE+\left.\f{\pd Y}{\pd L}\right|_{\bar{\EE},\bar{\LL}}\,\hat\delta\LL\, .
\ee
$\hat\delta\EE$ and $\hat\delta\LL$ are different from $\delta\EE,\delta\LL$ in the sense that the four-velocity is normalized with respect to either the background spacetime (hatted) or perturbed spacetime (no hat).
Thus, acting with $\hat\delta$ on Eq.~(\ref{eq:tdot_phidot}) we arrive at
\begin{align}
\hat{\delta}\dot{t} &= \f{(r^2+a^2)\,\hat{\delta}{\mathfrak T}}{r^2\Delta} +\f{a\,\hat{\delta}x}{r^2},\label{eq:delta_tdot}\\
\hat{\delta}\dot{\phi} &= \f{a\, \hat{\delta}{\mathfrak T}}{r^2\Delta}+\f{\hat{\delta}x}{r^2},\label{eq:delta_phidot}
\end{align}
where we used $\hat\delta r = 0$ and
\begin{align}
\hat{\delta}{\mathfrak T} &= \hat{\delta}\EE (r^2+a^2) -a\, \hat{\delta}\LL,\label{eq:delta_T}\\
\hat{\delta}x &= \hat{\delta}\LL - a\,\hat{\delta}\EE.\label{eq:delta_x}
\end{align}
This is consistent with our definition from Paper I. 
$\hat{\delta}\EE$ and $\hat{\delta}\LL$ can be thought of as $\ord(q)$ corrections to the energy and angular momentum due to the conservative part of the GSF.

The prescription for computing $\hat\delta\EE$ and $\hat\delta\LL$ is detailed in Sec.~III.D.1 of Paper I based on the work of Ref.~\cite{Barack:2011ed}. 
The computation requires the knowledge of the $t$ and $\phi$ components of the conservative GSF around a given eccentric orbit. \\
It is then straightforward to obtain 
$\hat{\delta}\dot{r}$ using $u_a u^a = -1 $ which yields
\be
\dot{r}\hat{\delta}\dot{r} = -\f{1}{\bar{g}_{rr}}\left[\bar{g}_{tt}\, \dot{t}\,\hat{\delta}\dot{t}+ \bar{g}_{t\phi}\,( \dot{t}\,\hat{\delta}\dot{\phi}+\dot{\phi}\,\hat{\delta}\dot{t})+\bar{g}_{\phi\phi}\, \dot{\phi}\,\hat{\delta}\dot{\phi}\right]\label{eq:delta_rdot}\, .
\ee
%
Next, using Eqs.~(\ref{eq:cco}) we can write the $\ord(q)$ coefficients $c_{0i}$ in the following compact form
$$ c_{0i} = \bar{g}_{ab}\, \delta u^a \bar{e}_i^b= \bar{g}_{ab} \left(\hat{\delta}u^a +\f{h_{00}}{2}\bar{u}^a\right)\bar{e}^b_i = \bar{g}_{ab}\, \hat\delta u^a \bar{e}_i^b,\qquad i =1,3\, ,$$
which then yields
\begin{align}
c_{01} &= \bar{g}_{tt}\, \bar{e}_1^t\,\hat{\delta}\dot{t}+ \bar{g}_{t\phi}\,( \bar{e}_1^t\,\hat{\delta}\dot{\phi}+\bar{e}_1^\phi\,\hat{\delta}\dot{t})+ \bar{g}_{rr}\, \bar{e}_1^r\,\hat{\delta}\dot{r}+
\bar{g}_{\phi\phi}\, \bar{e}_1^\phi\,\hat{\delta}\dot{\phi}, \label{eq:c01} \\ 
c_{03} &= \bar{g}_{tt}\, \bar{e}_3^t\,\hat{\delta}\dot{t}+ \bar{g}_{t\phi}\,( \bar{e}_3^t\,\hat{\delta}\dot{\phi}+\bar{e}_3^\phi\,\hat{\delta}\dot{t})+ \bar{g}_{rr}\, \bar{e}_3^r\,\hat{\delta}\dot{r}+
\bar{g}_{\phi\phi}\, \bar{e}_3^\phi\,\hat{\delta}\dot{\phi}. \label{eq:c03}
\end{align}
We are now ready to isolate the $\ord(q)$ correction to the geodetic spin precession.
Defining $\omega \equiv \omega_{[13]}= \f{1}{2}g_{ab}\left( e^a_3 \f{D e_1^b}{d\tau} - e^a_1 \f{D e_3^b}{d\tau}\right)$ as in Paper I, we obtain 
\be
\delta\omega = \f{1}{2}\bar\omega h_{00} + \delta\Gamma_{[31]0}+\left( c_{01} \bar{e}^b_1+c_{03}\bar{e}^b_3\right)\bar{e}_{a[3}\bar\nabla_b \bar{e}^a_{1]} + \f{1}{2}\f{d}{d\tau}(c_{13}-c_{31}), \label{eq:delta_omega}
\ee
where
\begin{eqnarray}
\delta \Gamma_{[31]0} &\equiv&  \tensor{\delta \Gamma}{^a _{cd}} \bar{e}_{a[3} \bar{e}_{1]}^c \bar{u}^d 
 = \left(\delta \Gamma_{a b c}  - h_{ad} \tensor{\bar\Gamma}{^d_{bc}} \right) \bar{e}^a_{[3} \bar{e}^b_{1]} \bar{u}^c ,
\end{eqnarray}
with $\delta \Gamma_{abc} \equiv \frac{1}{2} \left( h_{ab,c} + h_{ac,b} - h_{bc,a} \right)$ which has $3\times3\times4/2 = 18$ independent
components for equatorial Kerr. Now using Eq.~(\ref{eq:delta_Y_generic}) for $\Psi$ results in
\be
\delta\Psi = \int_0^{2\pi} \left(\f{\delta\omega}{\bar\omega}-\f{\delta\dot{r}}{\bar{u}^r} \right)\bar\omega\, \f{d\bar\tau}{d\chi}\,d\chi \label{eq:delta_Psi}.
\ee
%
%
Substituting $\delta\dot{r} = \hat\delta\dot{r} + \f{1}{2}h_{00}\bar{u}^r$ in this equation gives
\be
\delta\Psi = \int_0^{2\pi}\f{d\bar\tau}{d\chi} \left[\delta\Gamma_{[31]0}+\left( c_{01} \bar{e}^b_1+c_{03}\bar{e}^b_3\right)\bar{e}_{a[3}\bar\nabla_b \bar{e}^a_{1]}\right]d\chi 
+ \int_0^{2\pi}\left(-\f{\hat\delta\dot{r}}{\bar{u}^r}\right)\f{d\bar\Psi}{d\chi}\,d\chi \label{eq:delta_Psi_v2}.
\ee
%
%
The last term in Eq.~(\ref{eq:delta_omega}) integrates to zero over a radial orbit due to the periodicity of the motion in the perturbed spacetime
so it has been omitted. 
The evaluation of the background term $\bar{e}_{a[3}\bar\nabla_b\bar{e}^a_{1]} $
requires us to treat the tetrad as a field so that we may 
extend the covariant derivative off the worldline. We provided a justification of this extension in Sec.~III.C.2 of Paper I,
where we had also shown agreement for $\Delta\psi$ between a numerical GSF computation employing this extension and an analytic pN calculation independent of extension, hence validating the extension in Schwarzschild spacetime to the numerical accuracy of the GSF code.
Motivated by this, we extend the tetrad off the worldline for equatorial orbits in Kerr spacetime in the same manner.

Since $c_{01}$ and $c_{03}$ are both functions of  
$\hat\delta\{\dot{t},\dot{r},\dot{\phi}\}$ which in turn 
depend only on $\hat\delta\EE$ and $\hat\delta\LL$ at $\ord(q)$, we can rewrite Eq.~(\ref{eq:delta_Psi_v2}) as the sum of three separate integrals
\be
\delta\Psi = \int_0^{2\pi}\f{d\tau}{d\chi}\delta\Gamma_{[31]0}\,d\chi + \int_0^{2\pi} C_{\hat\delta\EE}\, \hat\delta\EE\, d\chi + \int_0^{2\pi} C_{\hat\delta\LL}\, \hat\delta\LL \,d\chi\label{eq:delta_Psi_v3}.
\ee
%
%
%

The first term in Eq.~(\ref{eq:delta_Psi_v3}) can be obtained in a straightforward fashion from the 
contraction of $\delta\Gamma_{abc}$ with
$\bar{e}^a_3\bar{e}^b_1\bar{u}^c$ given by Eqs.~(\ref{eq:e0} - \ref{eq:e3}). The resulting explicit expression is rather long
so we display in App.~\ref{sec:AppB}. $C_{\hat\delta\EE}$ and $C_{\hat\delta\LL}$ appear even more ungainly
when fully written out; hence they will not be displayed explicitly here. They can be computed without any hurdles.

Eqs.~(\ref{eq:Delta_Psi_formula}, \ref{eq:delta_Psi_v3}) are our master equations for $\Delta\psi$. In terms of computational strategy, the problem reduces
to determining the metric perturbation $h_{ab}$ and the GSF $F^a$ and using these to compute $\delta\Gamma_{abc}, \hat\delta\EE$, and $\hat\delta\LL$.

\section{The \texorpdfstring{$e\to0$}{} limits of \texorpdfstring{$\delta\omega$}{}, \texorpdfstring{$\delta\psi$}{} and \texorpdfstring{$\Delta\psi$}{} }\label{sec:e_to_limit}
We start by showing that in the $e\to 0$ limit, our expression for $\delta\omega$ as given by Eq.~(\ref{eq:delta_omega}) reduces to the circular-orbit result presented in Eq.~(2.65) of Dolan {\it et al.} 2014 \cite{Dolan:2014pja}, namely,
\be
\delta\omega_\c = \f{\bar\omega_\c}{2} \left(h_{00}-h_{11}-h_{33}\right) + \delta\Gamma_{\bar{3}\bar{0}\bar{1}} + \beta_{03} \bar{\Gamma}_{331} \label{eq:Dolan2014_265},
\ee
where $\bar\omega_\c=\bar\omega_{13}(e=0)=p^{-3/2};\;\beta_{03} = -\sqrt{p \Delta} F_r^\text{circ}/2$ contains the radial component of the circular-orbit GSF, $\delta\Gamma_{\bar{3}\bar{0}\bar{1}} = \left(h_{ac,b}+h_{ab,c}-h_{bc,a}\right) \eb_{3,\c}^a \bar{u}_\c^b \eb_{1,\c}^c/2$; and $\bar{\Gamma}_{331} =\bar\Gamma_{abc}\eb^a_{3,\c}\eb^b_{3,\c}\eb^c_{1,\c}$ is purely background. 
The semi-latus rectum $p$ reduces to the dimensionless circular-orbit radius, i.e., $p=r_0/m_2$.

First, we evaluate the $e\to 0$ limit of the orthonormal tetrad in the BL basis. Using $\dot{r}=0$, hence $\uu^r=\eb^t_1=\eb_1^\phi=\eb^r_3=0$, we obtain the circular tetrad straightforwardly with the caveat that $\eb_1^a$ is not unit normalized. After rescaling it with the appropriate factor, we obtain
\begin{align}
  \uu^a_\c& =\f{1}{p^{3/2}\, v\, \Omega_\text{circ}}[ 1, 0,0,\Omega_\text{circ}\,]^T , \label{eq:e0circ} \\
 \eb^a_{1,\c} &= \left[0,\f{\sqrt{\Delta}}{p} ,0,0\right]^T, \label{eq:e1circ} \\
 \eb^a_{3,\c} &= \left[\f{p^2-2a\sqrt{p}+a^2}{p^{3/2}\, v\, \Delta},0,0,\f{p^{3/2}-2\sqrt{p}+a}{p^{3/2}\, v\, \Delta}\right]^T \label{eq:e3circ},
\end{align}
where
\[ v = \sqrt{1-\f{3}{p}+\f{2a}{p^{3/2}}}, \qquad\text{and}\qquad \Omega_\c = \f{1}{p^{3/2}+a}\ .\]
It can be checked easily that the triad $\{  \uu_\c^a ,\eb_{1,\c}^a,\eb^a_{3,\c}\}$ is orthonormal and matches the circular-orbit one presented in Ref.~\cite{Dolan:2014pja}. 
For the remainder of this section it is understood that the metric perturbation along with all other quantities
are either evaluated at $e=0$ or at the limit $e\to 0$ so we mostly omit the sub/superscripts $\{\c,\e\}$ and display them as needed. 

We proceed by dropping the ambiguous last term in Eq.~(\ref{eq:delta_omega}) since it averages to zero over an eccentric orbit. Next, we look at the $\delta\Gamma_{[31]0}$ term which is explicitly shown in Eq.~(\ref{eq:deltaGamma310Antisym}).
Immediate substitution of the components of the tetrad into this equation yields
\be
\delta\Gamma_{[31]0}^\e =-\f{\Delta}{2 p^{7/2}} h_{rr} -\f{1}{2 p^{3/2}}h_{33}-\f{1}{2}\eb_{1}^r\left( \uu^a \eb_{3}^b h_{ra,b}\right) + \f{1}{2}\eb_{1}^r\left(\uu^a \eb_3^b h_{ab,r}\right)\, ,
\ee
where we introduced the superscript ${}^\e$ to denote $\Lim{\e}$.
Using the fact that the only nonzero component of $\eb_{1}^a$ is $\eb_1^r=\f{\sqrt{\Delta}}{p}$, we can turn the above expression into
\begin{align}
\delta\Gamma^\e_{[31]0} &=-\f{h_{11}}{2 p^{3/2}} -\f{h_{33}}{2 p^{3/2}}+\f{1}{2}\left( h_{ab,c} \eb_3^a \uu^b \eb_1^c -h_{ab,c} \eb_1^a \uu^b \eb_3^c \right),\nn \\
&=-\f{\bar\omega_\c}{2}(h_{11}+h_{33})+\f{1}{2}\left( h_{ab,c} -h_{cb,a} \right) \eb_3^a \uu^b \eb_1^c,\label{eq:delta_Gamma_310_Antisym_Circ_1}
\end{align}
where, for the very last term, we relabelled the indices so that it resembles $\delta\Gamma_{\bar{3}\bar{0}\bar{1}}$ defined above. The difference is given by
\begin{align}
 \f{1}{2}h_{ac,b} \eb^a_3 \uu^b \eb_1^c &= \f{1}{2} \eb_1^r \left[\eb^a_3\left(\uu^t h_{ar,t} + \uu^\phi h_{ar,\phi}\right)  \right] = 0\label{eq:zero_term_in_dG310}\, ,
\end{align}
since, for circular orbits, $h_{ab,t} = -\Omega_\c h_{ab,\phi} = -\f{\uu^\phi}{\uu^t} h_{ab,\phi}$. 
Adding Eq.~(\ref{eq:zero_term_in_dG310}) to (\ref{eq:delta_Gamma_310_Antisym_Circ_1}) we obtain
\be
\delta\Gamma^\e_{[31]0} =-\f{\bar\omega_\c}{2}(h_{11}+h_{33})+\delta\Gamma_{\bar{3}\bar{0}\bar{1}},\label{eq:delta_Gamma_310_Antisym_Circ_2}
\ee
resulting in
\begin{align}
 \delta\omega^\e &= \f{\bar\omega_\c}{2}\left(h_{00}-h_{11}-h_{33}\right) + \delta\Gamma_{\bar{3}\bar{0}\bar{1}} +\left[\left( c_{01} \eb^a_1+c_{03}\eb^a_3\right)\eb_{a[3}\bar\nabla_b \eb^a_{1]}\right]^\e \label{eq:delta_omega_circ_1} \ .
\end{align}
Next, we must show that the $c_{01}$ term disappears and the $c_{03}$ term goes to the $\beta_{03}$ term of Eq.~(\ref{eq:Dolan2014_265}). The coefficient multiplying $c_{01}$ is given by
$ \eb_1^r \eb_{a[3}\bar\nabla_r \eb_{1]}^a =0$,  since $\eb_1$ only has an $r$ component and $e_3^a$ does not. 
Now, we focus on the coefficient of the $c_{03}$ term. Writing it out explicitly we obtain
\begin{align}
 \f{1}{2}\left[\eb_{a3} \eb^b_3 \bar\pd_b \eb_1^a - \eb_{a1}\eb^b_3 \bar\pd_b \eb_3^a + \bar{\Gamma}_{331}-\bar{\Gamma}_{abc} \eb_1^a \eb_3^b \eb_3^c \right]^\e= \f{1}{2}\left[ \bar{\Gamma}_{331}-\left(-\bar{\Gamma}_{331}\right)\right]^\e = \bar{\Gamma}_{331}  \nn \ .
\end{align}
Putting these into Eq.~(\ref{eq:delta_omega_circ_2}) we arrive at
\be
\delta\omega^\e = \f{\bar\omega_\c}{2}\left(h_{00}-h_{11}-h_{33}\right) + \delta\Gamma_{\bar{3}\bar{0}\bar{1}}+c_{03}^\e\,\bar{\Gamma}_{331}  \label{eq:delta_omega_circ_2} \ .
\ee
The final step is to show that the $\e$ limit of $c_{03}$ reduces to $\beta_{03} =  -\sqrt{p \Delta} F_r^\text{circ}/2$. 
To this end, we combine techniques developed by Barack and Sago for the ISCO shift computation \cite{PhysRevD.81.084021} and for the original formulation of the eccentric redshift invariant \cite{Barack:2011ed}. First, returning to Eq.~(\ref{eq:c03}) with $\eb_3^r=0$, then using Eqs.~(\ref{eq:delta_tdot}) - (\ref{eq:delta_x}) we obtain
\be
c_{03}^\e = -\eb_{3,\c}^t\, \hat\delta E^\e + \eb_{3,\c}^\phi\, \hat\delta L^\e \label{eq:c03_circ_1} \ .
\ee
For orbits with $e\ll 1$ we can write $r(\tau) = r_0 (1-e\cos\omega_r \tau)$, 
where $\omega_r$ is the proper-time frequency of the radial oscillation about the circular orbit with radius $r_0 =m_2 p$. From Eq.~(3.28) of Paper I we have that
\be
\hat\delta E(\chi) = \hat\delta E(0)+\mathcal{E}(\chi), \quad  \hat\delta L(\chi) = \hat\delta L(0)+\mathcal{L}(\chi),\label{eq:hatdelta_E_L}
\ee
where
\be
\mathcal{E}(\chi) \equiv  -\int_0^\chi \f{d\bar\tau}{d\chi'} F_t^\text{cons}(\chi')\,d\chi', \qquad \mathcal{L}(\chi) \equiv \int_0^\chi \f{d\bar\tau}{d\chi'} F_\phi^\text{cons}(\chi')\,d\chi'\label{eq:calE_calL}\, 
\ee
and it is understood that the terms in the integrand also depend on $p$ and $e$ which we have suppressed. 
$\hat\delta E(0)$ and $ \hat\delta L(0)$ are shifts in the energy and angular momentum at the periapsis ($\chi=0$) due to the conservative effects of the GSF \cite{Barack:2011ed}. Since $F_t^\text{cons}$ and $F_\phi^\text{cons}$ scale as $\ord(e)$ as $\e$, $\mathcal{E}(\chi)$ and $\mathcal{L}(\chi)$ drop from
Eqs.~(\ref{eq:calE_calL}) and we end up with
\be
c_{03}^\e = -\eb_{3,\c}^t\, \hat\delta E(0)^\e + \eb_{3,\c}^\phi\, \hat\delta L(0)^\e \label{eq:c03_circ_2} \ .
\ee
An expression for $\hat\delta\{E(0),L(0)\}$ can be obtained by following the prescription detailed in Sec. II.C of Ref.~\cite{Barack:2011ed}. For Kerr spacetime,
we proceed by perturbing the $\dot{r}^2$ orbit equation
\be
\hat\delta \left[ \EE(r^2+a^2) - a \LL  \right]^2 = \Delta\,\hat\delta \left[ r^2+ (\LL-a \EE)^2\right] \label{eq:perturbed_orbit}
\ee 
keeping in mind that $\hat\delta r = 0$. 
Retaining the linear-in-$\hat\delta$ terms on both sides of Eq.~(\ref{eq:perturbed_orbit}) and evaluating the resulting expression at $\chi=\{0,\pi\}$ yields two equations
\begin{align}
 2\left[\bar\EE\left(r_\text{min}^2+a^2\right)-a\bar\LL\right]\left[\left(r_\text{min}^2+a^2\right)\Delhat E(0)-a\, \Delhat L(0)\right] &= 2 \Delta_\text{min} \left(\bar\EE a-\bar\LL\right) \left[a\, \Delhat E(0) - \Delhat L(0)\right],\nn\\
  2\left[\bar\EE\left(r_\text{max}^2+a^2\right)-a\bar\LL\right]\left[\left(r_\text{max}^2+a^2\right)\Delhat E(\pi)-a\, \Delhat L(\pi)\right] &= 2 \Delta_\text{max} \left(\bar\EE a-\bar\LL\right) \left[a\, \Delhat E(\pi) - \Delhat L(\pi)\right]\nn \, .
\end{align}
Substituting $\Delhat\EE(\pi) = \Delhat\EE(0)+\mathcal{E}(\pi)$, $\Delhat\LL(\pi) = \Delhat\LL(0)+\mathcal{L}(\pi)$ and $r_\text{min} = m_2 p/ (1+e)$, $r_\text{max}=m_2 p/(1-e)$ into the equations, solving for $\Delhat\{E(0),\LL(0)\}$ and finally keeping the leading-order term in the resulting small-$e$ expansions yields
\begin{align}
 \Delhat\EE(0)^\e &= \f{ \tilde\Delta\left[\Omega_\c \mathcal{L}(\pi)-\mathcal{E}(\pi)\right]^\e}{4 e\, p^{7/2} \,v^2\,\Omega_\c} \label{eq:DeltaE0},\\
 \Delhat\LL(0)^\e &= \f{ \tilde\Delta\left[\Omega_\c \mathcal{L}(\pi)-\mathcal{E}(\pi)\right]^\e}{4 e\, p^{7/2} \,v^2\,\Omega^2_\c} \label{eq:DeltaL0}\, ,
\end{align}
where $\tilde\Delta = p^2 - 2p +a^2$. Though it might at first seem alarming that $\Delhat\{E(0),\LL(0)\}$ scale as $\ord(e^{-1})$ in the $\e$ limit, we next
prove that $\left[\Omega_\c \mathcal{L}(\pi)-\mathcal{E}(\pi)\right]^\e \sim \ord(e)$ hence resulting in finite $\e$ limits. We insert Eqs.~(\ref{eq:DeltaE0}, \ref{eq:DeltaL0}) into our 
expression for $c_{03}^\e$ given in Eq.~(\ref{eq:c03_circ_2}) and obtain
\be
c_{03}^\e =\f{ \sqrt{\tilde\Delta}\left[\Omega_\c \mathcal{L}(\pi)-\mathcal{E}(\pi)\right]^\e}{4 e\, p^{2} \,v\,\Omega_\c} \label{eq:c03_circ_3}.
\ee
To evaluate this we use Ref.~\cite{PhysRevD.81.084021}'s argument that for $e\ll 1$ we can write
\begin{align}
 F_t &= F_t^\c + e\, \omega_r  F_t^\text{sin} \sin \omega_r \tau+\ord(e^2), \label{eq:Ft_expanded}\\
 F_\phi &= F_\phi^\c + e\, \omega_r  F_\phi^\text{sin} \sin \omega_r \tau+\ord(e^2)\label{eq:Fphi_expanded},\\
 F^r &=F^r_\c + e F_1^r \cos\omega_r\tau+\ord(e^2) \label{eq:Fr_expanded},
\end{align}
which we insert into the four-velocity relation $\uu^a F_a = 0$ and expand up to $\ord(e)$. 
Note that whereas $F^r_\c$ is a conservative-only piece, $F^\c_{t,\phi}$ are dissipative-only pieces for the circular-orbit case. Thus we can set them equal to zero  [$\uu^a F_a=0$ evaluated at $\ord(e^0)$ returns $F_t^\c + \Omega_\c F^\c_\phi=0$ regardless].
At $\ord(e)$ we obtain
\be 
\left. \uu^a F_a\right|_{\ord(e)} = \left. \uu^r \right|_{\ord(e)} F_r^\c + e\, \omega_r \sin(\omega_r\tau)\, \uu^t_\c\left[F_t^{\sin} + \Omega_\c F_\phi^{\sin}\right] + (\ldots) \cos\omega_r\tau = 0 , \label{eq:u_dot_F_expanded}
\ee
which gives
\be 
 e\, \omega_r \sin(\omega_r\tau)\, \uu^t_\c\left[F_t^{\sin} + \Omega_\c F_\phi^{\sin}\right]= -\left. \uu^r \right|_{\ord(e)} F_r^\c  + (\ldots) \cos\omega_r\tau , \label{eq:u_dot_F_expanded2}
\ee
where $|_{\ord(e)}$ denotes the $\ord(e)$ part only and we deliberately omitted the explicit expression for the coefficient of $\cos\omega_r\tau$ because it is only a function of $p$ so the orbit integral of the cosine term is oblivious to this coefficient. This will become more transparent in a few lines.
We are now in a position to evaluate $\left[\Omega_\c \mathcal{L}(\pi)-\mathcal{E}(\pi)\right]^\e$. Using Eqs.~(\ref{eq:calE_calL}) 
yields
\begin{align}
\left[\Omega_\c \mathcal{L}(\pi)-\mathcal{E}(\pi)\right]^\e&=e\int_0^\pi d\chi'\left[ \left.\f{d\bar\tau}{d\chi'}\right|^\e \, \omega_r \sin\omega_r\tau\,\left[F_t^{\sin} + \Omega_\c F_\phi^{\sin}\right]+(\ldots)\cos\omega_r\tau\right]\nn \\
&= -\int_0^\pi d\chi'  \left.\f{d\bar\tau}{d\chi'}\right|^\e\left( \f{\left. \uu^r\right|_{\ord(e)}}{\uu^t_\c}\, F^\c_r \right)\quad+\quad (\ldots)\int_0^\pi  \cos\chi'\,d\chi'\nn\\
&= -\int_0^\pi d\chi' \left.\f{dr}{d\chi'}\right|_{\ord(e)} \f{F_r^\c}{\uu^t_\c} \nn\\
&= -e \,p\, \f{F_r^\c}{\uu^t_\c} \int_0^\pi \sin\chi' d\chi'\nn\\ 
&= -2e\, p^{5/2} \, v\,\Omega_\c\, F_r^\c \label{eq:OmE_minusL}.
\end{align}
Above, going from the first line to the second we used Eq.~(\ref{eq:u_dot_F_expanded2}). 
The second integral in the second line is trivially zero. 
Finally, since $dr/d\chi$ starts at $\ord(e)$ we have 
$$\left.\uu^r\right|_{\ord{(e)}} = \f{\left.\f{dr}{d\chi}\right|_{\ord(e)}}{\left.\f{d\bar\tau}{d\chi}\right|^\e} .$$
Inserting the result of Eq.~(\ref{eq:OmE_minusL}) into Eq.~(\ref{eq:c03_circ_3}) we obtain
\be
c_{03}^\e = -\f{1}{2}\sqrt{p\tilde\Delta}\, F_r^\c = \beta_{03}\label{eq:c03_circ_4}.
\ee
To summarize, the $\e$ limit of our general expression for $\delta\omega$ written for eccentric equatorial orbits in Kerr spacetime reduces to
\be
\lim_\e\delta\omega = \f{\bar\omega_\c}{2}\left(h_{00}-h_{11}-h_{33}\right) + \delta\Gamma_{\bar{3}\bar{0}\bar{1}}+\beta_{03}\,\bar{\Gamma}_{331} =\delta\omega_\c \label{eq:delta_omega_circ_3}
\ee
in complete agreement with Ref.~\cite{Dolan:2014pja}. 

Going from $\delta\omega^\e$ to $\delta\psi^\e$ requires evaluation of a few more terms.
Using Eq.~(\ref{eq:delta_Psi}) in $\delta\psi^{e\to 0} = -\delta\Psi^{e\to 0}/\bar{\Phi}^{e\to 0}$ yields
\be
\delta\psi^{e\to 0} =\f{1}{\bar{\Phi}^{e \to 0}} {\left.\frac{d\bar\tau}{d\chi}\right|^{e\to 0}}\left[ \f{\bar{\omega}_\text{circ}}{2}( h_{11}+h_{33}) - \delta\Gamma_{\bar{3}\bar{0}\bar{1}}-\beta_{03}\bar{\Gamma}_{331}\right]+\lim_{e\to 0}\int_0^{2\pi}\left(\frac{-\hat\delta\dot{r}}{\bar{u}^r}\right)\f{d\bar\Psi}{d\chi}\, d\chi \, . \label{eq:delta_psi_eto0}
\ee
Using ${\left.\frac{d\bar\tau}{d\chi}\right|^{e\to 0}}(\bar{\Phi}^{e \to 0})^{-1}= v/\bar\omega_\text{circ}$, Eq.~(\ref{eq:c03_circ_4}) for $\beta_{03}$, and $\bar{\Gamma}_{331} = (p-1)/(p\sqrt{\tilde\Delta})$, we obtain
\be
\lim_\e \delta\psi = v \left[-\f{1}{\bar\omega_\text{circ}} \delta\Gamma_{\bar{3}\bar{0}\bar{1}} +\f{1}{2}(h_{11}+h_{33})+\f{1}{2}p(p-1)F_r^\text{circ}\right]
+\lim_{e\to 0}\int_0^{2\pi}\left(\frac{-\hat\delta\dot{r}}{\bar{u}^r}\right)\f{d\bar\Psi}{d\chi}\, d\chi \, . \label{eq:delta_psi_eto0_2}
\ee
In its current form, Eq.~(\ref{eq:delta_psi_eto0_2}) shows partial agreement with Eq.~(2.66) of Ref.~\cite{Dolan:2014pja}. To reach full analytic agreement, 
one must evaluate the last term containing $\hat\delta\dot{r}$ in the $e\to 0$ limit. However, from Eq.~(\ref{eq:delta_rdot}) we see that $\hat\delta\dot{r}/\bar{u}^r$ is manifestly $O(e^{-2})$ which 
means that we would have to extend the expansions in Eqs.~(\ref{eq:Ft_expanded}) and (\ref{eq:Fphi_expanded}) to  $\ord(e^2)$ $\big[ \Lim{e\to 0} {d}\bar\Psi/{d}\chi$ starts at $\ord(e^0)\big]$. This is precisely the same issue that we had encountered in Paper I when we evaluated $\Lim{\e}\delta\psi$ in Schwarzschild spacetime. For that case, we dealt with this issue
by computing the $\hat\delta\dot{r}$ term numerically for small, decreasing values of eccentricity and extrapolated these to $e\to 0$; hence showing the final
agreement $\delta\psi^{e\to 0} = \delta\psi_\text{circ}$ numerically. Likewise, here, we opt for the same resolution and delegate this numerical computation to upcoming work (Paper III). Thanks to the very recent work of Ref.~\cite{Kav_et_al}, the last term of Eq.~(\ref{eq:delta_psi_eto0_2}) can now be analytically computed in the Schwarzschild case, but the extension to the Kerr case is yet to come.

Recall that we wish to make a comparison between the circular-orbit result and the $e\to 0$ limit expression for the full gauge-invariant quantity $\Delta\psi$. In other words, we want to compare
\begin{align}
 \Delta \psi^{\text{circ}} &= \delta \psi^{\text{circ}} - \frac{d \bar{\psi}^{\text{circ}}}{d \bar{\Omega}^\c_\phi} \delta \Omega_\phi^\text{circ} 
 \end{align}
 with
 \begin{align}
\lim_{e \to 0} \Delta \psi &= \lim_{e \to 0} \delta \psi - \lim_{e \rightarrow 0} \left[ \frac{\partial \bar{\psi}}{\partial \bar{\Omega}_r} \delta \Omega_r +   \frac{\partial \bar{\psi}}{\partial \bar{\Omega}_\phi} \delta \Omega_\phi  \right] \label{eq:Delta_psi_e20_limit}.
\end{align}
As we have argued that $\Lim{e\to 0}\delta\psi = \delta\psi^\c$, we immediately obtain
\be
\lim_{e \to 0} \Delta \psi-\Delta\psi^\c = \left[  \frac{d \bar{\psi}^{\text{circ}}}{d \bar{\Omega}^\c_\phi}-\lim_{e\to 0} \frac{\partial \bar{\psi}}{\partial \bar{\Omega}_\phi}\right]\delta\Omega_\phi^\e- \lim_{e\to 0} \frac{\partial \bar{\psi}}{\partial \bar{\Omega}_r} \delta \Omega_r \, . \label{eq:Deltapsi_minus_DeltapsiCirc}
\ee
In Paper I we had shown that for Schwarzschild spacetime this term is given by a function of $p$ multiplied by $\Delta k$: the gauge-invariant $\ord(q)$ correction to the fractional periapsis advance in the $\e$ limit. Using
this definition and our notation from Paper I, we may write
\be
2\pi \Delta k = \Delta\Phi^\e = \delta\Phi^\e -\frac{d\bar\Phi^\c}{d\bar\Omega^\c_\phi}\,\delta\Omega_\phi^\e\equiv \bar{g}_1\,\delta\Omega_r^\e + \bar{g}_2\, \delta\Omega_\phi^\e\, \label{eq:delta_k},
\ee
where
\[
 \bar{g}_1 = -\f{1}{2\pi}\bar{T}^\e \bar\Phi^\e ,\quad\text{and}\quad \bar{g}_2 = \bar{T}^\e-\frac{d\bar\Phi^\c}{d\bar\Omega^\c_\phi}\, .
\]
The proportionality between Eqs.~(\ref{eq:Deltapsi_minus_DeltapsiCirc}) and (\ref{eq:delta_k}) implies that
\be
 \lim_{e \to 0} \Delta \psi-\Delta\psi^\c =
 \bar{G}\, \Delta k\,\label{eq:Deltapsi_minus_DeltapsiCirc2} ,
\ee
where
\be
\bar{G} =-\f{2\pi}{\bar{g}_1}\left( \lim_{e\to 0}\frac{\partial \bar{\psi}}{\partial \bar{\Omega}_r}\right)= \f{2\pi}{\bar{g}_2}\left(  \frac{d \bar{\psi}^{\text{circ}}}{d \bar{\Omega}^\c_\phi}-\lim_{e\to 0} \frac{\partial \bar{\psi}}{\partial \bar{\Omega}_\phi}\right)\, \label{eq:G}.
\ee
Note that the right-hand side of Eq.~(\ref{eq:G}) yields two distinct expression for $\bar{G}$; we numerically confirmed that these are equal. 
As they are rather long we do not display them here, but provide details in App. \ref{sec:AppA} for the interested reader. 
When $a=0$, $\bar{G}$ reduces to $-2\sqrt{p-3}\,(p-6)^{5/2}/\big[p\, (4p^2-39p+86)\big]$ in agreement with Eq.~(3.10) of Paper I. In Schwarzschild spacetime, the quantity $\Delta k$ is a gauge invariant in the usual sense used in the self-force
literature \cite{Barack:2011ed}. It has been computed using the GSF approach, pN theory and NR with consistent results \cite{LeTiec:2011bk} and has been analytically computed as a pN series up to $\ord(p^{-19/2})$ \cite{Bini:2016qtx}.

%
Eq.~(\ref{eq:Deltapsi_minus_DeltapsiCirc2}) is consistent with our conclusion from Paper I that the $e\to 0$ limit of $\Delta\psi$ does not agree with its circular-orbit counterpart $\Delta\psi^\c$. However, as $\Delta k$ is a gauge invariant in Schwarzschild spacetime, $\Lim{e\to 0}\Delta\psi$ is still a
gauge-invariant quantity as we explicitly demonstrated in Paper I via two independent approaches that agreed to high precision.  
Eq.~(\ref{eq:Deltapsi_minus_DeltapsiCirc2}) presents 
the Kerr result for this offset between the $e\to 0$ limit and the circular result. What needs to be established is whether or not $\Delta k=\Delta\Phi^\e/(2\pi)$, given by Eq.~(\ref{eq:delta_k}), truly is the gauge-invariant $\ord(q)$ correction to the periapsis advance
for equatorial orbits in Kerr spacetime, which was computed in a synergistic EOB-NR study \cite{PhysRevD.88.084005}. In Sec.~\ref{Sec:Formulation_GSF_method}, we provided the toolkit to compute $\Delta\Phi^\e$ [cf. Eqs.~(\ref{eq:delta_tdot_rdot_phidot}) --- (\ref{eq:delta_x})] via the $\ord(q)$ ingredients $h_{ab}$ and $F^a$ for which results were recently obtained along bound equatorial geodesics in Kerr spacetime \cite{vandeMeent:2015lxa, vandeMeent:2016pee}. 
%
\section{Discussion}\label{Sec:discussion} \vspace{0.1cm}
We have presented a formulation for computing the back-reaction induced correction to the geodetic spin precession for a spinning compact object in an eccentric, equatorial orbit
around a Kerr black hole. Denoted by $\Delta\psi$, this is a first-post-geodesic order  correction thus contributes at a relative magnitude of $\sim \ord(q)$ with respect to the background precession $\bar\psi$, where $q=m_1/m_2 \ll 1$. We have turned off the dissipation due to radiation reaction
for this computation; as such only the conservative parts of the 
metric perturbation $h_{ab}$ and the gravitational self-force $F^a$ contribute to $\Delta\psi$. Our formulation follows much
of the formalism laid out for the Schwarzschild case by Ref.~\cite{Akcay:2016dku} and leads us to similar results for eccentric, equatorial orbits in Kerr spacetime.
Our final expression for $\Delta\psi$ given by Eq.~(\ref{eq:Delta_psi1}) [with Eqs.~(\ref{eq:Delta_Psi_formula}, \ref{eq:delta_omega}), and (\ref{eq:delta_Psi_v2}) substituted]
is well suited for direct input from a self-force computation such as Ref.~\cite{vandeMeent:2016hel} which determined $h_{ab}$ and $F^a$ along eccentric, equatorial geodesics in Kerr spacetime.

Furthermore, we have proved, in detail, that the $e\to0$ limit of $\delta\omega$
fully agrees with the circular-orbit result of Ref.~\cite{Dolan:2014pja} (see Eq.~(\ref{eq:delta_omega_circ_3})). We have additionally shown 
that $\Lim{\e}\delta\psi$ agrees with $\delta\psi^\c$ modulo
one term which we argued should also agree based
on our work in Schwarzschild spacetime. With this caveat, we have then demonstrated that $\Lim{\e}\Delta\psi$ does not agree
with its circular counterpart $\Delta\psi^\c$.
This is a result that we had expected given that this non-equality also occurs in the Schwarzschild case. This
disagreement is due to the fact that we are comparing
$\Delta\psi^\c$, which is obtained by `fixing' the single orbital frequency $\Omega_\phi^\c$ (i.e., $\Delta\Omega_\phi^\c= 0$),
with the $e\to 0$ limit of $\Delta\psi$ which is obtained by
holding both $\Omega_r$ and $\Omega_\phi$ fixed. 
Because of this, the orbit in the perturbed spacetime will not necessarily be circular even when the background orbit is so.
We related the difference $\Lim{\e}\Delta\psi-\Delta\psi^\c$ to $\Delta k$:
the $\e$ limit of the gauge-invariant $\ord(q)$ correction to the general relativistic periapsis advance. Our result thus provides
a way to compute this quantity complementary to the EOB-NR
approach of Ref.~\cite{PhysRevD.88.084005}.

One may now wonder about the future directions based on this work.
As we have already mentioned, we are currently working on the next article (Paper III) which will supplement our work here with numerical results. More specifically, Paper III will (i) show the $\Lim{\e}\delta\psi = \delta\psi^\c$ agreement numerically; (ii) provide a data set for numerically computed $\Delta\psi$ that reasonably covers the $\{p,e,a\}$ parameter space of equatorial geodesics in Kerr spacetime; (iii) numerically compute the periapsis-advance correction $\Delta k$ and compare with existing results; (iv) compare the Schwarzschild $\Delta\psi$ (obtained in the Lorenz gauge)
with the $a=0$ Kerr $\Delta\psi$ obtained in radiation gauge; (v) use the high-precision radiation-gauge Kerr code to numerically extract the third- and possibly fourth order post-Newtonian contributions to $\Delta\psi$ in Schwarzschild spacetime.

Additionally, we will tackle the challenge of analytically deriving the next-order post-Newtonian (3-pN) contribution to $\Delta\psi$. However, this may take considerable effort and warrant a separate article. 
Ref.~\cite{Kav_et_al} recently obtained an analytical expression for $\Delta\psi$ based on Hopper {\it et al.}'s (\cite{Hopper:2015icj}) double-expansion in $p^{-1}$ and $e^2$, which goes up to $\ord(p^{-6})$ and $\ord(e^2)$. This agrees with the 2-pN expression of Paper I and will provide a check of our future pN results.

Finally, there remains the challenge of extending this work to completely generic (inclined and eccentric) orbits in Kerr spacetime. 
Ref. \cite{Bini:2017slb} recently obtained an expression for the precession frequency of the \emph{background} geodetic precession of a test gyroscope.
Similarly, Ref.~\cite{PhysRevD.95.084024}, motivated by naked singularities, derived an expression for the spin precession in any stationary spacetime.
More work still needs to be done in the unperturbed case in order for our linear-perturbation approach to yield a meaningful, gauge-invariant $\ord(q)$ correction. We leave the navigation of these uncharted waters to future work.

\acknowledgments
\vspace{0.1cm}
S.A.~acknowledges support from the Irish Research Council, funded under the National Development Plan for Ireland. S.A. also thanks Niels Warburton and Sam Dolan for a careful reading of this manuscript.

\vspace{0.3cm}
\appendix
\section{Zero-eccentricity limit of certain background quantities}\label{sec:AppA}
Below, we list the $e\to 0$ limits of the \emph{background} quantities of which we make use in Sec.~\ref{sec:e_to_limit}. 
As all quantities of interest are in the Kerr background we dispense with the overbars. 
Introducing $u\equiv p^{-1/2}$ and recalling that $v= \sqrt{1-3u^2+2 a u^3}, \tilde\Delta=p^2-2p+a^2$, we have
\begin{align}
\lim_\e\f{d\tau}{d\chi} &= \f{p^{3/2}\, v}{\sqrt{1-6u^2+8 a u^3-3a^2 u^4}},\label{eq:dtau_dchi_circ}\\
\lim_\e\f{d\Psi}{d\chi} &=  \f{ v}{\sqrt{1-6u^2+8 a u^3-3a^2 u^4}}\, .\label{eq:dPsi_dchi_circ}
\end{align}
Next, we list the $e\to 0$ limit of the partial derivative terms needed for Eq.~(\ref{eq:Delta_Psi_formula}):
\begin{align}
 \lim_{e\to 0}\frac{\pd\Psi}{\pd p} &= -\f{3\pi\, (\sqrt{p}-a)\,[p^2-2a\sqrt{p}(p-1)-a^2]}{p^{9/2}\, v\, (1-6u^2+8au^3-3a^2u^4)^{3/2}},\label{eq:dPsidp_eTo0}\\
 \lim_{e\to 0}\f{1}{e}\frac{\pd\Psi}{\pd e} &=3 \pi  (\sqrt{p}-a)^2\f{\left(\begin{array}{l}a^4 \sqrt{p} (13 p-81)+4 a^3 (47-12 p) p+a^2 p^{3/2} \left(-3 p^2+68 p-225\right)\\+2 a p^2 \left(p^2-22 p+72\right)-p^{5/2} \left(p^2-15 p+42\right)+14 a^5\end{array}\right)}{p^{11/2}\, v\,\tilde\Delta\,  (1-6u^2+8 a u^3-3a^2 u^4)^{5/2} }\, \label{eq:dPside_eTo0}.
\end{align}
Note that $\pd\Psi/\pd e$ starts at $\ord(e)$. As we show below the inverse Jacobian contains $\ord(e^{-1})$ terms that multiply this
hence returning an $\ord(e^0)$ result as desired. A quick check shows that Eqs.~(\ref{eq:dPsidp_eTo0}, \ref{eq:dPside_eTo0}) agree with their Schwarzschild limits
\be
\lim_{e\to 0}\left.\frac{\pd\Psi}{\pd p}\right|_{a=0}= -\f{3\pi}{\sqrt{p-3}\,(p-6)^{3/2}},\quad\text{and}\quad \lim_{e\to 0}\f{1}{e}\left.\frac{\pd\Psi}{\pd e}\right|_{a=0}=-\f{3\pi \, (p^2-15p+42)}{(p-2)\sqrt{p-3}\,(p-6)^{5/2}}\, .
\ee
We also need the $e\to 0$ limits of $\f{\pd\{p,e\}}{\pd \Omega_{\{r,\phi\}}}$ given by the inverse of the Jacobian $J^{-1}=\left[\f{\pd \Omega_{\{r,\phi\}}}{\pd\{p,e\}}\right]^{-1}$. To compute it, we start with
\begin{align}
 \lim_{e\to 0}\frac{\pd T}{\pd p} &= \f{\pi  \left(\begin{array}{l} 6 a^5 \sqrt{p}+a^4 \left(-3 p^2-2 p+8\right)-6 a^3 \sqrt{p} \left(p^2-p+6\right)+3 (p-8) (p-2)^2 p^3\\+2 a^2 p \left(-8 p^3+35 p^2-42 p+24\right)+6 a (p-2)^2 p^{3/2} (6 p-1)\end{array}\right)}{p^{7/2}\,(p-2)^2\, (1-6u^2+8au^3-3a^2u^4)^{3/2}},\label{eq:dTdp_eTo0}\\
 \lim_{e\to 0}\f{1}{e}\frac{\pd T}{\pd e} &=\f{\pi   \left(\begin{array}{l}-3 a^9 \left(9 p^2-32 p+4\right)+a^8 \sqrt{p} \left(9 p^3+152 p^2-540 p+160\right)+2 a^7 p \left(27 p^3-431 p^2+966 p-456\right)\\
 +2 a^6 p^{3/2} \left(41 p^4-455 p^3+2050 p^2-3056 p+1344\right)\\
 +a^5 p^2 \left(-353 p^4+3052 p^3-10349 p^2+13108
   p-4484\right)\\
   +a^4 p^{5/2} \left(31 p^5+4 p^4-1817 p^3+8948 p^2-13172 p+4288\right)\\
   -2 a^3 p^3 \left(160 p^5-1627 p^4+5559 p^3-7018 p^2+1716 p+1080\right)\\
   +2 a^2 p^{7/2} \left(-18 p^6+579 p^5-4527 p^4+14602 p^3-20996 p^2+11016 p+192\right)\\
   +3 a (p-2)^3 p^4
   \left(38 p^3-376 p^2+841 p-2\right)+3 (p-2)^3 p^{11/2} \left(2 p^3-32 p^2+165 p-266\right)\end{array}\right)}{p^5\,(p-2)^3\, \tilde\Delta\,(1-6u^2+8 a u^3-3a^2 u^4)^{5/2}}\, ,\label{dTde_eTo0}\\
  \lim_{e\to 0}\frac{\pd \Phi}{\pd p} &= \f{2 \pi  \left[3 a^4 \sqrt{p}+a^3 (4-10 p)+a^2 \sqrt{p} \left(-5 p^2+23 p-18\right)+6 a (p-2)^2 p-3 (p-2)^2 p^{3/2}\right]}{p^{7/2}\,(p-2)^2\, (1-6u^2+8au^3-3a^2u^4)^{3/2}},\label{eq:dPhidp_eTo0}\\
  \lim_{e\to 0}\f{1}{e}\frac{\pd \Phi}{\pd e} &=\f{\pi  \left(\begin{array}{l}-3 a^8 \left(9 p^2-32 p+4\right)+8 a^7 \sqrt{p} \left(25 p^2-84 p+20\right)+2 a^6 p \left(38 p^3-467 p^2+1186 p-456\right)\\-32 a^5 p^{3/2} \left(12 p^3-81 p^2+164 p-84\right)+a^4 p^2 \left(-23 p^4+848 p^3-4213 p^2+7524 p-4484\right)
  \\+8 a^3 p^{5/2}   \left(7 p^4-116 p^3+490 p^2-840 p+536\right)\\
  +2 a^2 p^3 \left(p^5-21 p^4+233 p^3-950 p^2+1660 p-1080\right)-48 a (p-2)^3 p^{7/2}+3 (p-2)^4 p^4\end{array}\right)}{p^5\,(p-2)^3\,\tilde\Delta\,(1-6u^2+8 a u^3-3a^2 u^4)^{5/2}}\, ,\label{dPhide_eTo0}
\end{align}
The $a\to0$ limits of these agree with their Schwarzschild expressions,
\begin{align}
\lim_{e\to 0}\left.\frac{\pd T}{\pd p}\right|_{a=0}&= \f{3\pi p\, (p-8)}{(p-6)^{3/2}},\\ 
\lim_{e\to 0}\f{1}{e}\left.\frac{\pd T}{\pd e}\right|_{a=0}&=\frac{3 \pi  \, p^2 \left(2 p^3-32 p^2+165 p-266\right)}{(p-2)(p-6)^{5/2} },\\
\lim_{e\to 0}\left.\frac{\pd \Phi}{\pd p}\right|_{a=0}&= -\frac{6 \pi }{\sqrt{p}\,(p-6)^{3/2} },\\ 
\lim_{e\to 0}\f{1}{e}\left.\frac{\pd \Phi}{\pd e}\right|_{a=0}&=\frac{3 \pi  \sqrt{p}}{(p-6)^{5/2}}\, .
\end{align}
From the above expressions, we can now obtain the Jacobian $\f{\pd \Omega_{\{r,\phi\}}}{\pd\{p,e\}}$ and its inverse $\f{\pd \{p,e\}}{\pd\Omega_{\{r,\phi\}}}$
in a straightforward fashion. Then using
\be
\f{\pd \Psi}{\pd\Omega_i} = \int_0^{2\pi} \left[ \f{\pd p}{\pd\Omega_i}\f{\pd}{\pd p}\f{\pd\Psi}{\pd\chi}+\f{\pd e}{\pd\Omega_i}\f{\pd}{\pd e}\f{\pd\Psi}{\pd\chi}\right]d\chi\,
\ee
and recalling that $\Omega_\c = (p^{3/2}+a)^{-1}$ we finally obtain
\begin{align}
 \lim_{e\to 0}& \frac{\pd\Psi}{\pd \Omega_r}= \f{6\pi (a-\sqrt{p})\tilde\Delta^2 \, v}{\Omega_\text{circ}\, p\,(1-6u^2+8 a u^3-3a^2 u^4) }\nn\\   &\times{\left(\begin{array}{l}
                                                 3 a^6 (66-29 p) \sqrt{p}+a^5 p \left(18 p^2+185 p-458\right)+a^4 p^{3/2} \left(-124 p^2+69 p+214\right)\\
                                                  +a^3 p^2 \left(-49 p^3+504 p^2-1148 p+864\right)+a^2 p^{5/2} \left(149 p^3-1004 p^2+2172 p-1584\right)\\
                                                  +3 a (p-2)^2 p^3 \left(4 p^2-53   p+82\right)-3 (p-2)^2 p^{7/2} \left(2 p^2-15 p+14\right)+ 15 a^7 (p-2)
                                                 \end{array}\right)}\nn \\
                                                 \times&{\left(\begin{array}{l}
                                                 9 a^{10} \left(p^2-16 p+4\right)-36 a^9 \sqrt{p} \left(3 p^2-34 p+8\right)+3 a^8 p \left(-21 p^3+248 p^2-1576 p+504\right)\\
                                                 -4 a^7 p^{3/2} \left(3 p^3+371 p^2-2560 p+1436\right)-3 a^6 p^2 \left(73 p^4-906 p^3+1731 p^2+2812 p-4036\right)\\
                                                 +12 a^5 p^{5/2} \left(159   p^4-1329 p^3+3290 p^2-2172 p-472\right)                                          +a^4 p^3 (299 p^5-7744 p^4+47585 p^3\\-115086 p^2+112956 p-32424)
                                                 -12 a^3 (p-2)^2 p^{7/2} \left(99 p^3-1005 p^2+2600 p-1716\right)\\
                                                 -3 a^2 (p-2)^2 p^4 \left(22 p^4-634 p^3+4337 p^2-10188 p+7620\right)\\
                                                 +36 a   (p-2)^4 p^{9/2} \left(2 p^2-31 p+90\right)-9 (p-2)^4 p^5 \left(4 p^2-39 p+86\right)
                                                 \end{array}\right)^{-1}}\,\label{eq:dPsidOmega_r_eTo0}\, , \\
 \lim_{e\to 0}&\frac{\pd\Psi}{\pd \Omega_\phi}=  \f{6\pi (a-\sqrt{p})\tilde\Delta }{\Omega_\text{circ}\, p^{5}\,v\,(1-6u^2+8 a u^3-3a^2 u^4)^{3/2} }\nn\\   &\times{\left(\begin{array}{l}
 (p-2) \sqrt{p} \left(\sqrt{p}-a\right) \Big[-14 a^5+a^4 (81-13 p) \sqrt{p}+4 a^3 p (12 p-47)+a^2 p^{3/2} \left(3 p^2-68 p+225\right)\\
 -2 a p^2 \left(p^2-22 p+72\right)+p^{5/2} \left(p^2-15 p+42\right)\Big] \Big[6 a^5 \sqrt{p}+a^4 \left(-3 p^2-2 p+8\right) -6   a^3 \sqrt{p} \left(p^2-p+6\right)\\
 +2 a^2 p \left(-8 p^3+35 p^2-42 p+24\right) +6 a (p-2)^2 p^{3/2} (6 p-1)+3 (p-8) (p-2)^2 p^3\Big]\\
 +\left(a^2+2 a (p-1) \sqrt{p}-p^2\right) \Big[-3 a^9 \left(9 p^2-32 p+4\right) +a^8 \sqrt{p} \left(9 p^3+152 p^2-540   p+160\right)\\
 +2 a^7 p \left(27 p^3-431 p^2+966 p-456\right) +2 a^6 p^{3/2} \left(41 p^4-455 p^3+2050 p^2-3056 p+1344\right)\\
 +a^5 p^2 \left(-353 p^4+3052 p^3-10349 p^2+13108 p-4484\right)\\
 +a^4 p^{5/2} \left(31 p^5+4 p^4-1817 p^3+8948 p^2-13172 p+4288\right)\\
 -2a^3 p^3 \left(160 p^5-1627 p^4+5559 p^3-7018 p^2+1716 p+1080\right)\\
 +2 a^2 p^{7/2} \left(-18 p^6+579 p^5-4527 p^4+14602 p^3-20996 p^2+11016 p+192\right)\\
 +3 a (p-2)^3 p^4 \left(38 p^3-376 p^2+841 p-2\right)+3 (p-2)^3 p^{11/2} \left(2 p^3-32 p^2+165   p-266\right)\Big]
 \end{array}\right)}\nn \\
                                                 \times&{\left(\begin{array}{l}
                                                 -9 a^{10} \left(p^2-16 p+4\right)+36 a^9 \sqrt{p} \left(3 p^2-34 p+8\right)+3 a^8 p \left(21 p^3-248 p^2+1576 p-504\right)\\
                                                 +4 a^7 p^{3/2} \left(3 p^3+371 p^2-2560 p+1436\right)+3 a^6 p^2 \left(73 p^4-906 p^3+1731 p^2+2812p-4036\right)\\
                                                 -12 a^5 p^{5/2} \left(159   p^4-1329 p^3+3290 p^2-2172 p-472\right)\\
                                                 +a^4 p^3 \left(-299 p^5+7744 p^4-47585 p^3+115086 p^2-112956 p+32424\right)\\
                                                 +12 a^3 (p-2)^2 p^{7/2} \left(99 p^3-1005 p^2+2600 p-1716\right)\\
                                                 +3 a^2 (p-2)^2 p^4 \left(22 p^4-634 p^3+4337 p^2-10188 p+7620\right)\\
                                                 -36 a   (p-2)^4 p^{9/2} \left(2 p^2-31 p+90\right)+9 (p-2)^4 p^5 \left(4 p^2-39 p+86\right)
                                                 \end{array}\right)^{-1}}\,\label{eq:dPsidOmega_phi_eTo0}.
\end{align}
Though these may appear ungainly their $a\to 0$ limits agree with their Schwarzschild counterparts which are
\begin{align}
 \lim_{e\to 0}\left.\frac{\pd\Psi}{\pd \Omega_r}\right|_{a=0} &=-\frac{2 \pi  p^2\,\sqrt{p-3}  \left(2 p^2-15 p+14\right)}{(p-6) \left(4 p^2-39 p+86\right)} \, \\
 \lim_{e\to 0}\left.\frac{\pd\Psi}{\pd \Omega_\phi}\right|_{a=0} &=\frac{2 \pi  p^{3/2}\, (p-2) \left(2 p^2-17 p+28\right)}{\sqrt{(p-3)(p-6)} \left(4 p^2-39 p+86\right)}\, .
\end{align}
From the above expressions, it is trivial to obtain $\Lim{e\to 0}\f{\pd\psi}{\pd\Omega_i}$ using
\be
\f{\pd\psi}{\pd\Omega_i}= -\f{1}{\Phi}\f{\pd\Psi}{\pd\Omega_i}+\f{\Psi}{\Phi^2}\f{\pd\Phi}{\pd\Omega_i}.
\ee
The resulting expressions are given by
\begin{align}
 \lim_{e\to 0}\frac{\pd\psi}{\pd \Omega_r}&= -\f{1}{\Phi^\e}\,(\ref{eq:dPsidOmega_r_eTo0}) -\f{v}{\Omega_\text{circ}\sqrt{1-6u^2+8 a u^3-3a^2 u^4}}\label{eq:dpsidOmega_r_eTo0}, \\
 \lim_{e\to 0}\frac{\pd\psi}{\pd \Omega_\phi}&=-\f{1}{\Phi^\e}\,(\ref{eq:dPsidOmega_phi_eTo0})+\f{p\,(p-2)\,v}{\Omega_\text{circ}\, \tilde\Delta}\label{eq:dpsidOmega_phi_eTo0}\,,
\end{align}
where $ \Phi^\e = 2\pi \tilde\Delta/\Big[ p\,(p-2)\,\sqrt{1-6u^2+8 a u^3-3a^2 u^4}\,\Big]$. 
Eqs.~(\ref{eq:dpsidOmega_r_eTo0}, \ref{eq:dpsidOmega_phi_eTo0}) once again agree with the Schwarzschild expressions in the $a=0$ case:
\begin{align}
 \lim_{e\to 0}\left.\frac{\pd\psi}{\pd \Omega_r}\right|_{a=0} &= -\frac{2\, p^{3/2}\, \sqrt{p-3}\,(p-6)^{3/2} }{4p^2-39p+84}\, \\
 \lim_{e\to 0}\left.\frac{\pd\psi}{\pd \Omega_\phi}\right|_{a=0} &=\,\frac{p \left(2 p^3-30 p^2+141 p-202\right)}{\sqrt{p-3} \left(4 p^2-39 p+86\right)} .
\end{align}
On the other hand, for the circular, equatorial Kerr case we have
\begin{align}
\f{\pd \Psi^\text{circ}}{\pd \Omega^\text{circ}_\phi} &= \f{2\pi \big[a^3+a^2 \sqrt{p} (2 p-3)+a (2-3 p) p+p^{5/2}\big]}{\Omega^2_\text{circ}\, p^5 \, v\, (1-6u^2+8 a u^3-3a^2 u^4)^{3/2} }\, \label{eq:dPsidOmegacirc},\\
\f{\pd \psi^\text{circ}}{\pd \Omega^\text{circ}_\phi} &= \f{a^3 (5 p-2)+a^2 \sqrt{p} \left(4 p^2-13 p+6\right)-3 a (p-2)^2 p+3 (p-2)^2 p^{3/2}}{3\Omega_\text{circ}^2\, v\, p^2 \tilde\Delta^2}\, ,\label{eq:dpsidOmegacirc}
\end{align}
which, respectively, agree with the corresponding Schwarzschild expressions $\frac{2 \pi  p^{5/2}}{(p-6)^{3/2} \sqrt{p-3}}$ and $\f{p}{\sqrt{p-3}}$ when $a=0$.
From these, we can clearly see the disagreement between $\Lim{e\to 0}\pd \{\Psi,\psi\}/\pd \Omega_\phi$ and $\pd\{\Psi^\c, \psi^\c\}/\pd \Omega^\c_\phi$.

\section{Explicit expression for \texorpdfstring{$\delta\Gamma_{[31]0}$}{} for bound equatorial orbits in Kerr spacetime}\label{sec:AppB}
In this section, we dispense with the overbar notation as the only $\ord(q)$ quantities are the various components of $h_{ab}$ hence all other quantities are background. Making a note of the factor of two on the left-hand side below, we have
\begin{align}
2\delta\Gamma_{[31]0}&= \frac{(h_{\text{r$\phi $}} e_3^r+h_{\text{t$\phi $}} e_3^t+h_{\phi \phi } e_3^{\phi })
   \left(a^2 (e_1^r u^{\phi }+e_1^{\phi } u^r)-a (e_1^r u^t+e_1^t u^r)-(r-2) r^2 (e_1^r u^{\phi }+e_1^{\phi }
   u^r)\right)}{r^2 \left(a^2+(r-2) r\right)}\nn\\
   &+\frac{(h_{\text{r$\phi $}} e_1^r+h_{\text{t$\phi $}}
   e_1^t+h_{\phi \phi } e_1^{\phi }) \left(a^2 (-(e_3^r u^{\phi }+e_3^{\phi } u^r))+a (e_3^r u^t+e_3^t
   u^r)+(r-2) r^2 (e_3^r u^{\phi }+e_3^{\phi } u^r)\right)}{r^2 \left(a^2+(r-2) r\right)}\nn   \\
   -(h_{\text{rr}}   e_3^r&+h_{\text{r$\phi $}} e_3^{\phi }+h_{\text{tr}} e_3^t)\nn\\
   \times&\left(\frac{u^t \left(a^2+(r-2) r\right)
   (e_1^t-a e_1^{\phi })}{r^4}+\frac{u^{\phi } \left(a^2+(r-2) r\right) \left(a^2 e_1^{\phi }-a e_1^t+r^3
   (-e_1^{\phi })\right)}{r^4}+\frac{e_1^r u^r \left(a^2-r\right)}{r \left(a^2+(r-2)
   r\right)}\right)\nn\\
   +(h_{\text{rr}} e_1^r&+h_{\text{r$\phi $}} e_1^{\phi }+h_{\text{tr}} e_1^t) \nn\\
   \times&\left(\frac{u^t
   \left(a^2+(r-2) r\right) (e_3^t-a e_3^{\phi })}{r^4}+\frac{u^{\phi } \left(a^2+(r-2) r\right) \left(a^2
   e_3^{\phi }-a e_3^t+r^3 (-e_3^{\phi })\right)}{r^4}+\frac{e_3^r u^r \left(a^2-r\right)}{r \left(a^2+(r-2)
   r\right)}\right)\nn\\
   &+\frac{(h_{\text{tr}} e_3^r+h_{\text{tt}} e_3^t+h_{\text{t$\phi $}} e_3^{\phi }) \left(a^3
   (e_1^r u^{\phi }+e_1^{\phi } u^r)-a^2 (e_1^r u^t+e_1^t u^r)+3 a r^2 (e_1^r u^{\phi }+e_1^{\phi } u^r)-r^2
   (e_1^r u^t+e_1^t u^r)\right)}{r^2 \left(a^2+(r-2) r\right)}\nn\\
   &-\frac{(h_{\text{tr}} e_1^r+h_{\text{tt}}
   e_1^t+h_{\text{t$\phi $}} e_1^{\phi }) \left(a^3 (e_3^r u^{\phi }+e_3^{\phi } u^r)-a^2 (e_3^r u^t+e_3^t
   u^r)+3 a r^2 (e_3^r u^{\phi }+e_3^{\phi } u^r)-r^2 (e_3^r u^t+e_3^t u^r)\right)}{r^2 \left(a^2+(r-2)
   r\right)}\nn\\
   &+\partial _rh_{\text{r$\phi $}} e_1^r e_3^{\phi } u^r-\partial _rh_{\text{r$\phi $}} e_1^{\phi }
   e_3^r u^r+\partial _rh_{\text{tr}} e_1^r e_3^t u^r-\partial _rh_{\text{tr}} e_1^t e_3^r u^r+\partial
   _rh_{\text{tt}} e_1^r e_3^t u^t-\partial _rh_{\text{tt}} e_1^t e_3^r u^t+\partial _rh_{\text{t$\phi $}}
   e_1^r e_3^t u^{\phi }\nn\\
   &+\partial _rh_{\text{t$\phi $}} e_1^r e_3^{\phi } u^t 
   -\partial _rh_{\text{t$\phi $}}   e_1^t e_3^r u^{\phi }
   -\partial _rh_{\text{t$\phi $}} e_1^{\phi } e_3^r u^t+\partial _rh_{\phi \phi } e_1^r
   e_3^{\phi } u^{\phi }-\partial _rh_{\phi \phi } e_1^{\phi } e_3^r u^{\phi }-\partial _th_{\text{rr}} e_1^r
   e_3^t u^r+\partial _th_{\text{rr}} e_1^t e_3^r u^r \nn\\
   &-\partial _th_{\text{r$\phi $}} e_1^r e_3^t u^{\phi   }
   +\partial _th_{\text{r$\phi $}} e_1^t e_3^r u^{\phi }+\partial _th_{\text{r$\phi $}} e_1^t e_3^{\phi }
   u^r-\partial _th_{\text{r$\phi $}} e_1^{\phi } e_3^t u^r-\partial _th_{\text{tr}} e_1^r e_3^t u^t+\partial
   _th_{\text{tr}} e_1^t e_3^r u^t+\partial _th_{\text{t$\phi $}} e_1^t e_3^{\phi } u^t\nn\\
   &-\partial   _th_{\text{t$\phi $}} e_1^{\phi } e_3^t u^t+\partial _th_{\phi \phi } e_1^t e_3^{\phi } u^{\phi }
   -\partial_th_{\phi \phi } e_1^{\phi } e_3^t u^{\phi }+\partial _{\phi }h_{\text{rr}} u^r (e_1^{\phi } e_3^r-e_1^r
   e_3^{\phi })+\partial _{\phi }h_{\text{r$\phi $}} u^{\phi } (e_1^{\phi } e_3^r-e_1^r e_3^{\phi })-\partial
   _{\phi }h_{\text{tr}} e_1^r e_3^{\phi } u^t\nn\\
   &-\partial _{\phi }h_{\text{tr}} e_1^t e_3^{\phi } u^r+\partial_{\phi }h_{\text{tr}} e_1^{\phi } e_3^r u^t
   +\partial _{\phi }h_{\text{tr}} e_1^{\phi } e_3^t u^r-\partial
   _{\phi }h_{\text{tt}} e_1^t e_3^{\phi } u^t+\partial _{\phi }h_{\text{tt}} e_1^{\phi } e_3^t u^t-\partial
   _{\phi }h_{\text{t$\phi $}} e_1^t e_3^{\phi } u^{\phi }+\partial _{\phi }h_{\text{t$\phi $}} e_1^{\phi }
   e_3^t u^{\phi }  \label{eq:deltaGamma310Antisym} \ .
\end{align}

\bibliographystyle{apsrev4-1}
\bibliography{references}

\end{document}